# Spintronics Detection of Interfacial Magnetic Switching in a Paramagnetic Tris(8-hydroxyquinoline)iron(III) Thin Film

Dali Sun[1], Christopher M. Kareis[2], Kipp J. van Schooten[1], Wei Jiang[3], Gene Siegel[3], Marzieh Kavand[1], Royce A. Davidson[2], William W. Shum[2], Chuang Zhang[1], Ashutosh Tiwari[3], Christoph Boehme[1], Feng Liu[3], Peter W. Stephens,[4] Joel S. Miller[2], Z. Valy Vardeny[1],*

[1]Department of Physics & Astronomy, University of Utah, Salt Lake City, Utah, 84112.

[2]Department of Chemistry, University of Utah, Salt Lake City, Utah, 84112

[3]Department of Material Science & Engineering, University of Utah, Salt Lake City, Utah 84112.

[4]Department of Physics & Astronomy, Stony Brook University, Stony Brook, NY 11794.

Organic semiconductors find increasing importance in spin transport devices due to the modulation and control of their properties through chemical synthetic versatility. The organic materials are used as interlayers between two ferromagnet (FM) electrodes in organic spin valves (OSV), as well as for magnetic spin manipulation of metal-organic complexes at the molecular level. In the latter, specifically, the substrate-induced magnetic switching in a paramagnetic molecule has been evoked extensively, but studied by delicate surface spectroscopies. Here we present evidence of the substantial magnetic switching in a nanosized thin film of the paramagnetic molecule, tris(8-hydroxyquinoline)iron(III) (Feq$_3$) deposited on a FM substrate, using the magnetoresistance response of electrical 'spin-injection' in an OSV structure; and the inverse-spin-Hall effect induced by state-of-art pulsed microwave 'spin-pumping'. We show that interfacial spin control at the molecular level may lead to a macroscopic organic spin transport device; thus, bridging the gap between organic spintronics and molecular spintronics.

*To whom correspondence should be addressed: val@physics.utah.edu

# I. INTRODUCTION

Organic semiconductors (OSEC) have attracted intense attention for potential applications in spintronic-based devices [1,2] because of the long spin relaxation time obtained for spin ½ carriers [3]. To date organic spintronics research has focused on the physics of the spin injection and spin transport through the organic interlayer in OSV devices. Detection of spin transport through the OSEC layer has been done through a variety of techniques that include magneto-transport [3-12], inverse spin Hall effect [13,14], muon spin rotation (µSR) [15,16], and two-photon photoemission [17,18]. In most applications the spin control in the device has been achieved via the injected spin-aligned carriers from conventional FM electrodes into the OSEC interlayer, in spite of the conductivity mismatch at their interface that poses a formidable barrier for spin injection [19].

In contrast to organic spintronics, 'molecular spintronics' utilizes the chemical versatility of molecules; in particular those that have paramagnetic metal ions, for manipulating the spin states [20-28]. One particularly promising class of building blocks for molecular spintronics devices is the metalloporphyrins, which exhibit an intrinsic remnant magnetization when in contact with a FM metallic electrode [23], similar to single molecule magnets [29]. Recently, metallophthalocyanines (e.g. CuPc [27], MnPc [30]) also have been intensely studied due to their potential highly spin polarized surface spins ('spinterface') that can act as a spin filter. However, the spin orientation of the molecular ensemble, which is crucial to the ability of spin filtering, was only investigated in the limit of monolayer using a variety of surface science techniques [23-31].

Here we report a spin current-based detection scheme of a molecular spin ensemble by incorporating the paramagnetic semiconductor tris(8-hydroxyquinoline)iron [32] (Feq$_3$; shown in Figure. 1a and S. I. Fig. S1-S3) as an interlayer into two macroscopic spintronic devices: (i) a FM/OSEC/Feq$_3$/Au device configuration (OSV-like device) for magnetoresistance response from electrically injected spin aligned carriers; and (ii) a FM/OSEC/Feq$_3$/Pt device configuration for inverse spin Hall effect (ISHE-type device) response from microwave (MW) pumped pure spin currents. In an OSV-like device, the Feq$_3$ layer functions as a spin filter; but, surprisingly it also exhibits a 'switching field' that mimics the coercive field of a conventional FM film. Consequently, the device magnetoresistance response, MR($B$) is similar to that of a more conventional OSV device. Using SQUID magnetometry we verified the substantial magnetic ordering and switching that occur in the Feq$_3$ layer, which is attributed to an indirect antiferromagnetic (AFM) exchange interaction with the FM metallic electrode in the device. In an ISHE-type device, due to this AFM exchange, the NiFe/Feq$_3$ layer generates a pure spin current having an *opposite* direction of spin polarization to the magnetization of NiFe. This device configuration results in ISHE response from an underlying Pt electrode to exhibit reverse polarity that contrasts to that of a NiFe/Pt device without the Feq$_3$ spacer. Our experimental findings are further supported by first-principles DFT-type calculations.

## II. EXPERIMENTAL DETAILS

Compared to the more conventional diamagnetic tris(8-hydroxyquinoline)aluminum (Alq$_3$), which has been widely used as OSEC interlayer in OSV devices [4], Feq$_3$ has *five electron spins* that originate from the 3d transition metal Fe$^{III}$ ion (Fig. 1a inset and S. I. Fig. S1); so that the ground state spin quantum number is S=5/2 [33]. In addition, the Feq$_3$ film is an air-stable semiconductor with energy gap and resulting photoluminescence emission that peaks at ~1.65 eV (~750 nm) (Fig. 1a); when measured without an attached FM layer it exhibits paramagnetic Curie-Weiss susceptibility behavior ($\chi \propto 1/T$) with no detectable hysteresis (Fig. 1b).

A schematic structure of the OSV-like device based on a Feq$_3$ interlayer is illustrated in Fig. 1c. The device consists of a bottom FM metallic electrode, an Alq$_3$ overlayer, Feq$_3$ interlayer, and capped with a nonmagnetic Au top electrode; a magnetic field, ***B*** is applied parallel to the device substrate. Due to the structural similarity of Feq$_3$ and Alq$_3$ molecules, these two organic layers maintain excellent interfacial contact in the device. Importantly, the inter-diffusion problem that occurs when a top FM metal electrode is evaporated onto a relatively soft organic layer at high deposition temperature (required in a conventional OSV) [4-9], is circumvented in this type of device (Fig. 1d) by replacing the top FM electrode with the Feq$_3$ organic spin filter layer that requires a much lower thermal deposition temperature. For an ISHE-type device, the Au electrode is replaced by Pt metal which is widely used to act as the spin sink for the detection of spin current due to its large spin-orbit coupling. In this case the whole configuration of layers is reversed upside down (namely Pt/Alq$_3$/Feq$_3$/FM).

We fabricated the OSV-type devices on two types of bottom FM electrodes. One is half-metal FM La$_{0.67}$Sr$_{0.33}$MnO$_3$ thin film that was epitaxially grown on SrTiO$_3$ substrates by pulsed laser deposition; and fabricated for bottom electrode using conventional wet-etch optical lithography. Another is the Ni$_{80}$Fe$_{20}$ bottom electrode that was grown by e-beam evaporation through a shadow mask on Si$_3$N$_4$ (400 nm)/Si substrates in a vacuum chamber devoted for metal deposition. The FM electrodes were subsequently transferred without breaking the vacuum into a second chamber devoted to OSEC deposition. The Alq$_3$ (Aldrich) and Feq$_3$ (synthesized by literature method [32]) films were grown *in situ* by thermal evaporation. The fabricated structures were transferred back to the metal deposition chamber for e-beam evaporation of an Au top electrode (25 nm) in a crossbar configuration. Typical device area was ~ 200 × 500 μm.

For an ISHE-type device, an Al thin film electrode (150 nm) was firstly grown on a glass template (3×50 mm) by sputtering using conventional optical lithography. Subsequently two Cu contacts (30 nm thick) with a gap of 3 mm (extended from an Al bottom electrode) were grown by e-beam evaporation through a shadow mask, followed by a strip of Pt electrode (3.5 mm × 1 mm × 7 nm). Without breaking the vacuum, the fabricated structures were transferred with another shadow mask to the organic deposition chamber for OSEC deposition. The OSEC deposition was similar to that used for the OSV-like device. Then ferromagnetic layer (Ni$_{80}$Fe$_{20}$, 15 nm), SiO$_2$ (500 nm) dielectric layer and top Cu thin film (30 nm) were all grown in series on the OSEC layer by e-beam evaporation through a shadow mask on the OSEC materials.

Transport measurements performed using a Quantum Design Physical Property Measurement System (PPMS-9) combined with a Keithley 2400 source meter. The magnetic field, *B*, was applied parallel to the device substrate. The MR is defined as: MR($B$) = ($R(B)$ - $R(0)$)/$R(0)$, where $R(0)$ is the junction resistance at $B = 0$, and $R(B)$ is the resistance measured at field $B$ using the four-points method. The magnetization measurements for the susceptibility and devices were performed using the Quantum Design MPMS-5 5 T superconducting quantum interference device (SQUID) magnetometer. The p-ISHE measurements were carried out at room temperature in a Bruker ElexSys E580 X-band (~9.7 GHz) pulsed EPR spectrometer equipped with a dielectric resonator (Bruker FlexLine ER 4118 X-MD5). The MW pulse duration time was set to 2 μs at a repetition rate of 500 Hz. The maximum pulsed MW power was ~1 kW resulting in an excitation field amplitude $B_1$=1.1 mT at the sample location. The p-ISHE(B) response measurements and time dynamics required averaging over 10240 shots. First-principles calculations were carried out using local spin density approximation (LSDA) with onsite Coulomb interactions and projector augmented-wave method in Vienna ab-initio simulation package (VASP) based on density functional theory, in which an additional on-site Hubbard-U term is included on the iron(III) (U=6.0 eV, J=0.9 eV).

## III. RESULTS AND DISCUSSION

### A. Magnetoresistance measurements

Typical MR(B) responses of various OSV-like devices are presented in Figs. 2a-2d. We note that the NiFe-based OSV-like device (Fig. 2a) was cooled to 5 K in a magnetic field of +300 mT. A jump of ~0.2% is observed in the MR(B) response when the NiFe magnetization switches at the FM coercive field, $B_{C1} \approx 3$ mT. This is confirmed by the anisotropic MR(B) response of the NiFe electrode in a OSV-like device (S. I. Fig. S4). The maximum MR(B) value, $MR_{max}$, obtained in this OSV-like device is comparable to NiFe-based conventional OSV devices [15,16]. We found, surprisingly that the MR(B) response switches back to the low resistance state at $B=B_{C2}$ ~25 mT, showing a similar response to MR(B) observed in conventional OSV, although only one FM electrode is used in the device configuration here. This indicates that an unusual magnetic ordering occurs in the Feq$_3$ layer when it is placed in the OSV-like device configuration, which is induced by the external field in combination with the bottom FM metallic electrode. As a control experiment, upon replacing the bottom NiFe electrode by an Au electrode to form an Au/Feq$_3$/Au diode, no MR response was obtained (S. I. Fig. S5). Thus, any explanation for the MR(B) response in the OSV-like device based on an intrinsic MR effect, such as organic MR (OMAR) [34] in the Feq$_3$ layer, may be excluded. Other explanations such as δB mechanism [35] can be also ruled out because of the absence of fringe field from the NiFe electrode, which exhibits uniform in-plane magnetization (confirmed by SQUID measurement below in Fig. 3a).

When we replaced the bottom NiFe electrode in the OSV-like device by the half-metal FM (see S. I. Fig. S6 for *I-V* characteristics), LSMO that has ~100% spin aligned carrier injection capability [36], then MR$_{max}$ is enhanced by an order of magnitude to ~2.7% (Fig. 2b), and the switching field, $B_{C2}$ increases to 50 mT. We note that in order to obtain the MR response in the LSMO-based OSV-like device, the non-hysteresis linear MR($B$) response that originates from the LSMO electrode [37] was subtracted from the measured MR($B$) (S. I. Fig. S7). The larger MR$_{max}$ observed for the LSMO-based OSV-like device indicates that spin aligned carrier injection indeed occurs into the OSEC interlayer, consistent with the different abilities of NiFe and LSMO FM electrodes to inject spin aligned carriers into an OSEC (40% and higher MR have been reported when using LSMO-based conventional OSV [4-12], whereas 0.1 to 0.3% MR have been reached when using NiFe-based OSV [15,16]). As a comparison, the MR($B$) response of a LSMO/Feq$_3$(50 nm)/Au device, where the Feq$_3$ layer is in direct contact with the LSMO electrode was also measured. This device shows an even larger MR$_{max}$~5.4%, with similar MR($B$) response and $B_{C2}$ value as that of the LSMO-based OSV-like device (Fig. 2c). The obtained MR$_{max}$ of LSMO-based 'OSV-like' devices based on different Alq$_3$ interlayer thickness, $d$ is summarized in Fig. 2d; MR$_{max}$ decreases monotonically with $d$. Importantly, the Feq$_3$ switching field $B_{C2}$ also monotonically decreases with $d$ (S. I. Fig. S8). The changes of $B_{C2}$ with $d$ provides strong evidence that the observed MR($\boldsymbol{B}$) response does not originate from 'tunneling anisotropy magnetoresistance' (TAMR) [38], in which the switching field is determined by the anisotropy magnetization of the bottom FM electrode, and remains unchanged at different $d$'s.

We also measured the MR($B$) response in both NiFe and LSMO-based OSV-like devices at different temperatures, $T$. MR$_{max}$ vs. $T$ for these three devices is summarized in Fig. 2e (see also Fig. S9). Similar to conventional OSV devices [4], MR$_{max}$ (and with Alq$_3$ spacer, Fig. S9) decreases steeply with increasing $T$ and vanishes at 100 K (130K) for NiFe (LSMO)-based OSV-like devices. In contrast, the MR response in the NiFe/Feq$_3$/Au device survives up to 200 K, indicating that the presumed magnetic ordering of the Feq$_3$ in the proximity with the bottom FM layer in fact persists to higher temperature. This implies that the temperature dependent spin diffusion length in the Alq$_3$ layer mostly limits the MR response for the OSV-like devices, consistent with the reported temperature dependence of Alq$_3$ spin diffusion length measured by the μSR technique [15]. The Coulomb blockade induced magnetoresistance cannot explain our observations either, since it occurs at very low temperature (below 1 K) [39]. Note, however that MR$_{max}$($d$) in LSMO/Feq$_3$ based devices decreases slightly faster than that in LSMO/Alq$_3$/Feq$_3$ based devices. This indicates that the LSMO/Alq$_3$ interface provides an additional spin filtering effect for the spin injection [9-11].

Taking together the various MR($B$) and MR$_{max}$ responses vs. thickness, temperature and voltage (S. I. Fig. S10), we thus infer that the OSV-like devices based on Feq$_3$ layer instead of a second FM behave very similar to conventional OSV devices that contain two FM electrodes. Therefore we conclude that the OSV-like device may be considered as a simplified version of OSV, which is based on a single FM electrode [40,41].

Remarkably, the spin filtering effect from the Feq$_3$ layer exhibits a "spin memory" behavior that leads to a polarity change in the MR($B$) response in the OSV-like devices, which depends on the magnetic field cooling history (Fig. 2f). When the NiFe/Alq$_3$/Feq$_3$/Au device is cooled to cryogenic temperatures subjected to a 'positive' $B$=+300 mT, then the device shows a transition from a "low resistance state" to a "high resistance state" that occurs at the NiFe $B_C$; this is a positive MR response (Fig. 2f, upper panel). In contrast, when the same device is cooled under the influence of a 'negative' $B$= -1 T, the device shows a reverse MR($B$) polarity of a 'negative' MR response (Fig. 2f, lower panel). The MR polarity reversal was also observed in NiFe/Feq$_3$/Au devices, but is not clear in the LSMO-based OSV-like devices (S. I. Fig. S11).

**B. Magnetization measurements**

To elucidate the apparent FM-like behavior of the paramagnetic Feq$_3$ layer, magnetization measurements (namely $M(B)$) were performed on all fabricated OSV-like devices. At variance with the previously reported FM ordering in metalloporphyrins and metallophthalocyanines monolayer detected by delicate surface science techniques [23-25], a substantial FM ordering of the Feq$_3$ layer in the OSV-like device configuration was observed using conventional magnetometry 'SQUID' measurements (Fig. 3). First we checked that the $M(B)$ response of a pristine NiFe film (Fig. 3a) shows an abrupt hysteretic response at $B < 2$ mT, consistent with its coercive field, $B_{C1}$. Next we checked the magnetization response of Feq$_3$ based structures. Compared to the linear paramagnetic response of pristine Feq$_3$ film having S=5/2 in the ground state (Fig. 1b), the $M(B)$ loops of NiFe/Feq$_3$, NiFe/Alq$_3$/Feq$_3$ and LSMO/Alq$_3$/Feq$_3$ 'OSV-like film structures' clearly show a second hysteretic transition at a higher field (Figs. 3b to 3d). This is distinct from the abrupt transition of the NiFe (or LSMO) electrode seen at low field. In addition, the narrow hysteresis response of the NiFe electrode at ~$B_{C1}$ is broader in the 'OSV-like film structures' than that of the pristine NiFe film. This magnetic 'hardening' originates from the OSEC overlayer, and is consistent with the enhanced exchange interaction found previously for π-conjugated molecules deposited on FM surfaces due to the proximity of the molecules to the FM atoms [42].

The $M(B)$ responses of NiFe/Feq$_3$ and NiFe/Alq$_3$/Feq$_3$ 'OSV-like film structures' measured upon cooling under two different and opposite magnetic fields of +300 mT and -1 T, are shown respectively in Figs. 3e and 3f. The $M(B)$ response asymmetry with respect to $B = 0$ is seen when the field is swept in one direction and then to the opposite direction. This indicates the presence of a 'magnetic exchange bias' [43,44], which results from an AFM coupling [23] at the interface between the Feq$_3$ and NiFe layers. We note that π-conjugated nonmagnetic organic molecules deposited on FM metallic film show only a symmetric $M(B)$ response [4,12,42]. Re-orientated easy axis on the surface of NiFe/Feq$_3$ layer from in-plane to out-of-plane can be ruled out because the total magnetization along the in-plane direction is unchanged in opposite field cooling. Surprisingly, the AFM coupling between the FM and Feq$_3$ layers still occurs, although much weaker, in the NiFe/Alq$_3$/Feq$_3$ 'OSV-like structure' even when the Feq$_3$ layer and NiFe film are separated by ~15 nm thick layer of the diamagnetic Alq$_3$ (Fig. 3f). This implies that the AFM interaction between the FM substrate and Feq$_3$ layer may be mediated by the Alq$_3$ layer via the hydroxyquinoline ligands [23]. In any case, the SQUID magnetometry measurements unambiguously reveal that the paramagnetic Feq$_3$ layer in the proximity of the FM substrate is *magnetically ordered*, consistent with the observed MR($B$)-type response of the OSV-like devices. The resulting magnetic switching of the remnant field in the Feq$_3$ layer occurs at $B_{C2} \gg B_{C1}$, and this generates the OSV-like MR($B$) response in the OSV-like devices.

## C. Inverse Spin-Hall effect measurements

A formidable known barrier for electrical 'spin injection' is the conductivity mismatch between FM and OSEC materials [19]. In contrast, the spin current generated via microwave-induced magnonic 'spin-pumping' overcomes the conductivity mismatch [14,45], and offers a perfect alternative to the 'spin injection' approach for the FM/OSEC interface. Furthermore, the inverse spin Hall effect (ISHE) measurement technique may provide a reliable detection scheme for the spin accumulation through the OSEC layers, without interference from many artifacts known to exist in the electrical spin-injection method (such as TAMR, TMR, AMR, etc.).

Fig. 4a demonstrates the working principle and schematic structure of an ISHE device based on Feq$_3$ molecules. The magnetization dynamics ***M(t)*** under ferromagnetic resonance (FMR) condition induces a pure spin current (***J$_S$***) in the adjacent non-magnetic Pt layer via spin pumping. Since Pt has a large spin Hall angle ($\theta_{SH} \sim 0.06$) [46], therefore the spin current leads to a related electric field, ***E$_{ISHE}$*** perpendicular to ***J$_S$*** and spin polarization ***S***: namely $\boldsymbol{E_{ISHE}} = \theta_{SH} \boldsymbol{J_S} \times \boldsymbol{S}$. We have used a state-of-the-art pulsed MW excitation [46] to deliver high MW power (~ 1 kW) to the FM substrate that consequently generates high spin current density in the Pt layer with minimum thermal/resonant heating effect (see S. I. Fig. S12 and Ref. 47). With the pulsed ISHE (p-ISHE) method we can investigate a Spinterface feature that occurs in Feq$_3$ layer only several molecular monolayers thick.

The inset of Fig. 4b shows the p-ISHE voltage generated from a NiFe/Pt ISHE device without Feq$_3$, measured at room temperature with an in-plane (i.e. $\theta_B$=0º) external magnetic field, ***B***, as illustrated in Fig. 4a. The p-ISHE response (***V*****$_{ISHE}$** ~ -1.3 mV at $\theta_B$=0º) is about two orders of magnitude larger than that of the cw-ISHE due to the high pulsed MW excitation intensity [46]. Possible heating effect can be excluded because its magnetic field response is independent of the ***B*** direction, in sharp contrast with the symmetric p-ISHE response between $\theta_B$=0º and $\theta_B$=180º in Fig. 4a [46,48]. When a 7 nm thick Feq$_3$ layer (~7 monolayers) is inserted in between the NiFe and Pt layers, the observed p-ISHE response from the Pt layer is reduced to ~ 76 µV (see S. I. Fig. S13 for MW power dependence of p-ISHE responses). Importantly, the p-ISHE polarity at $\theta_B$=0º is *reversed* (black line in Fig. 4c). The decrease in p-ISHE response cannot be simply attributed to loss of spin current in the Feq$_3$ layer ($\propto \exp(-d/\lambda)$, where $\lambda$ is the spin diffusion length), because the reduction in the ISHE by more than an order of magnitude is too large for a 7 nm Feq$_3$ layer with $\lambda \sim d$. Also, the p-ISHE polarity would remain unchanged, namely same as in the inset of Fig. 4b for NiFe/Pt structure, if the spin current would be directly generated from the NiFe layer into the Pt layer via pinholes. We thus conclude that the observed p-ISHE response in the NiFe/Feq3/Pt structure originates from the spin current that is pumped into the Pt layer from the Feq3 layer itself; we note that spin-pumping from a paramagnetic layer was recently demonstrated [49].

Due to the AFM exchange interaction between NiFe and Feq3 layers inferred from our MR(*B*) measurement and the SQUID measurements, the induced Feq3 magnetization, *m* is opposite to *M*. Consequently *m* in the Feq3 layer precesses under the influence of the dynamic magnetization *M*(t) in the NiFe layer, in the opposite direction, thereby generating magnons with opposite spin **S** respect to those in the NiFe layer. The generated magnons, in turn produce spin current at the Feq3/Pt interface having opposite spin direction to that produced without the Feq3 layer, and therefore $E_{ISHE}$ in the Pt layer reverses polarity (see right panel in Fig. 4a). We note, in passing that the electron paramagnetic resonance for the paramagnetic Feq3 molecules measured at the MW frequency that we use here (~9.7 GHz) is ~300 mT ($g \approx 2$), which is far away from the obtained FMR in the NiFe layer. We also measured the p-ISHE responses in a trilayers with smaller Feq3 thickness (~5 nm) (FIG. S14). The p-ISHE polarity in this device is still reversed as compared to the NiFe/Pt device. In addition the p-ISHE response is indeed larger (~93 µV) due to the enhanced exchange coupling at smaller Feq3 thickness. When a 15 nm thick Alq3 layer (same thickness as in the OSV-like device) is inserted in between the NiFe and Feq3 layers (Fig. 4d), the reversed pISHE response in the Pt layer decreases by about half in NiFe/Alq3/Feq3/Pt device, but still can be observed. This further confirms that the observed p-ISHE response in the Pt layer is directly related to the Feq3 layer rather than the diffused spin current that passes through the Alq3 layer [15]. We therefore conclude that the ISHE method provides direct evidence for a robust AFM exchange interaction between the NiFe and Feq3 layer, which is consistent with our observations of MR(*B*) and SQUID measurements in the OSV-like device.

### D. DFT calculations for the interaction between the FM substrate and Feq3 film

We carried out density functional theory (DFT) electronic-structure calculations for the Feq3 molecules in intimate contact with a FM substrate. To deduce the magnetic ordering strength within the Feq3 layer in proximity to the FM substrate, we extract the exchange coupling constant ($J_{ex}$) among the Feq3 molecules by fitting the Heisenberg spin lattice model to the DFT-calculated energy difference between FM and AFM states (see S.I. section 11). For a free-standing Feq3 molecular monolayer (Fig. 5a), the energy difference, $\Delta E$ between AFM and FM spin configuration is very small ($\Delta E = E_{AFM} - E_{FM} < 0.1$ meV, the energy convergence criteria is set at 0.1 meV), which translates to a negligible $J_{ex}$ (~ 0.002 meV) ; this indicates a paramagnetic free-standing Feq3 layer (S. I. Fig. S15). However, when the Feq3 layer is deposited onto the FM NiFe substrate that forms interface layer (Fig. 5b), $\Delta E$ between AFM and FM spin configuration becomes much larger ($\Delta E = E_{AFM} - E_{FM}$ ~ 8 meV), and the effective coupling among the Feq3 molecules changes to strong FM coupling with $J_{ex}$ ~ 0.8 meV. This indicates that the paramagnetic Feq3 layer transitions to weak FM ordering (S. I. Fig. S16a), similar to the Fe-porphyrin layer [23,25]. The origin of FM ordering in the Feq3 sublayer A can be understood from analysis of the spin-resolved, partial density of state (p-DOS) of the NiFe/Feq3 system (S. I. Fig. S16b). When the Feq3 molecules are in close proximity with the NiFe substrate, although there is no direct overlap of $Fe^{III}$ and NiFe $d$-orbitals, there exist Fe-O, Ni-O and Fe-N, Ni-N interactions, as deduced from the spin DOS (S. I. Fig. S16b), which are able to mediate a 'super-exchange'-like interaction. Furthermore, the results of first-principles calculations indicate that interface Feq3 layer prefers an AFM interface-mediated coupling with the underlying FM NiFe substrate, with an energy difference, $\Delta E = (E_{FM} - E_{AFM}) > 25$ meV (S. I. Fig. S16a). This is in agreement with the observed exchange bias in the obtained $M(B)$ response and reversed p-ISHE response (Figs. 3 and 4).

We also carried out similar calculations with an Alq3 layer in between the Feq3 layer and NiFe electrode (S. I. Fig. S17). The substrate-induced FM magnetic coupling within the Feq3 layer still exists, but is much weaker ($J_{ex}$ ~ 0.1 meV in this case). This is consistent with the experimental results of $MR_{max}$ and $B_{C2}$ as a function of the Alq3 buffer layer thickness presented in Fig. 2d and S. I. Fig. S8, respectively.

### III. SUMMARY

Our discovery of the versatile spin filter functionality of Feq3 thin films and its ability to form an OSV-like device is an important advance for organic spintronics applications. We demonstrate two spin-current based detection themes for studying the emergent magnetic response in molecules containing paramagnetic metal ions in the vicinity of a FM substrate. We show that both the OSV-like MR($B$) and reversed ISHE($B$) response originate from the magnetic ordering that occurs at the Feq$_3$/FM interface. Using a variety of chemical synthesis techniques, incorporation of different transition metals (e.g. Mnq$_3$, Crq$_3$, etc.) and other ligands or a proper FM substrate should enable tuning of the FM/OSEC exchange coupling, as well as the degree of magnetic ordering at the molecular level for altering the magnitude of MR($B$), ISHE and magnetization responses at the macroscopic level. Additionally, the spin filtering effect from the Feq$_3$ layer exhibits a remarkable "spin memory" behavior that leads to a polarity change in the MR($B$) response in the OSV-like devices, which depends on the magnetic field cooling history. This 'spin memory' behavior provides another dimension in controlling the spin response, enabling the development of more complex multi-functional organic spintronics devices.

**References:**

1. V. A. Dediu, L. E. Hueso, I. Bergenti, and C. Taliani, Spin routes in organic semiconductors. *Nat. Mater.* **8,** 707–716 (2009).
2. D. Sun, E. Ehrenfreund, and Z. V. Vardeny, The first decade of organic spintronics research. *Chem. Comm.* **50**, 1781-1793 (2013).
3. S. Pramanik, et al. Observation of extremely long spin relaxation times in an organic nanowire spin valve. *Nat. Nanotech.* **2**, 216-219 (2007).
4. Z. H. Xiong, D. Wu, Z. V. Vardeny, and J. Shi, Giant magnetoresistance in organic spin-valves. *Nature* **427**, 821–824 (2004).
5. J. W. Yoo et al. Spin injection/detection using an organic-based magnetic semiconductor. *Nat. Mater.* **9**, 638-642 (2010).
6. T. D. Nguyen et al. Isotope effect in spin response of π-conjugated polymer films and devices. *Nat. Mater*. **9**, 345–352 (2010).
7. D. Sun, et al. Giant Magnetoresistance in organic spin valves. *Phys*. *Rev*. *Lett*. **104**, 236602 (2010).
8. T. S. Santos et al. Room-Temperature Tunnel magnetoresistance and spin-polarized tunneling through an organic semiconductor barrier. *Phys*. *Rev*. *Lett*. **98**, 16601 (2007).
9. C. Barraud, et al. Unraveling the role of the interface for spin injection into organic semiconductors. *Nat*. *Phys*. **6**, 615–620 (2010).
10. S. Sanvito, Molecular spintronics: The rise of spinterface science. *Nat*. *Phys*. **6**, 562–564 (2010).
11. M. Galbiati, Spinterface: Crafting spintronics at the molecular scale. *MRS* Bulletin, **39**, 602 (2014)
12. D. Sun, et al, Active control of magnetoresistance of organic spin valves using ferroelectricity. *Nat*. *Commun*. **5**, 4396 (2014).
13. K. Ando, S. Watanabe, S. Mooser, E. Saitoh and H. Sirringhaus, Solution-processed organic spin-charge converter. *Nat*. *Mater*. **12**, 622-627 (2013).
14. S. Watanabe, et al. Polaron spin current transport in organic semiconductors. *Nat*. *Phys*. **10**, 308-313 (2014).

15. A. J. Drew, et al. Direct measurement of the electronic spin diffusion length in a fully functional organic spin valve by low-energy muon spin rotation. *Nat. Mater.* **8**, 109–114 (2009).
16. L. Schulz et al. Engineering spin propagation across a hybrid organic/inorganic interface using a polar layer. *Nat. Mater.* **10**, 39–44 (2011).
17. M. Cinchetti et al. Determination of spin injection and transport in a ferromagnet/organic semiconductor heterojunction by two-photon photoemission. *Nat. Mater.* **8**, 115–119 (2009).
18. S. Steil et al. Spin-dependent trapping of electrons at spinterfaces. *Nat. Phys.* **9**, 242-247 (2013).
19. G. Schmidt, D. Ferrand, L. W. Molenkamp, A. T. Filip and B. J. van Wees, Fundamental obstacle for electrical spin injection from a ferromagnetic metal into a diffusive semiconductor. *Phys. Rev. B* **62**, R4790 (2000).
20. A. R. Rocha et al. Towards molecular spintronics. *Nat. Mater.* **4**, 335-339 (2005).
21. S. Sanvito, Molecular spintronics. *Chem. Soc. Rev.* **40**, 3336–3355 (2011).
22. S. Sanvito, Organic spintronics: Filtering spins with molecules. *Nat. Mater.* **10**, 484-485 (2011)
23. H. Wende et al. Substrate-induced magnetic ordering and switching of iron porphyrin molecules. *Nat. Mater.* **6**, 516-520 (2007).
24. P. Gambardella et al. Supramolecular control of the magnetic anisotropy in two-dimensional high-spin Fe arrays at a metal interface. *Nat. Mater.* **8**, 189-193 (2009).
25. M. Mannini et al. Magnetic memory of a single-molecule quantum magnet wired to a gold surface. *Nat. Mater.* **8**, 194-197 (2009).
26. A. L. Rizzini et al. Coupling single molecule magnets to ferromagnetic substrates. *Phys. Rev. Lett.* **107**, 177205 (2011)
27. M. Warner et al. Potential for spin-based information processing in a thin-film molecular semiconductor. *Nature* **503**, 504-508 (2013).
28. T. Miyamachi et al. Robust spin crossover and memristance across a single molecule. *Nat. Commun.* **3**, 938, (2012).
29. L. Bogani and W. Wernsdorfer, Molecular spintronics using single-molecule magnets. *Nat. Mater.* **7**, 179-186 (2008).
30. F. Djeghloul et al. Direct observation of a highly spin-polarized organic spinterface at room temperature. *Sci. Rep.* **3**, 1272 (2013).

31. S. Schmaus et al. Giant magnetoresistance through a single molecule. *Nat. Nanotech.* **6**, 185-189 (2011).

32. J. E. Tackett and D. T. Sawyer, Properties and infrared spectra in the potassium bromide region of 8-quinolinol and its metal chelates. *Inorg. Chem.* **3**, 692-696 (1964).

33. W. Jiang et al. Structural, electronic, and magnetic properties of tris(8-hydroxyquinoline)iron(III) molecules and its magnetic coupling with ferromagnetic surface: first-principles study. J. Phys.: Condens. Matter. **28**, 176004 (2016).

34. Ö. Mermer et al. Large magnetoresistance in nonmagnetic π-conjugated semiconductor thin film devices. *Phys. Rev. B* **72**, 205202 (2005).

35. M. Cox, S. P. Kersten, J. M. Veerhoek, P. Bobbert, and B. Koopmans. ΔB mechanism for fringe-field organic magnetoresistance. *Phys. Rev. B* **91**, 165205 (2015).

36. J.-H. Park, et al. Direct evidence for a half-metallic ferromagnet. *Nature* 392, 794–796 (1998).

37. D. Wu, Z. Xiong, X. Li, Z. V. Vardeny and J. Shi, Magnetic-field-dependent carrier injection at $La_{2/3}Sr_{1/3}MnO_3$ and organic semiconductors interfaces. *Phys. Rev. Lett.* **95**, 016802 (2005)

38. M. Grünewald et al. Tunneling anisotropic magnetoresistance in organic spin valves. *Phys. Rev. B* **84**, 125208 (2011).

39. M. Urdampilleta, S. Klyatskaya, J. P. Cleuziou, M. Ruben and W. Wernsdorfer, Supramolecular spin valves. *Nat. Mater.* **10**, 502-506 (2011).

40. K. V. Raman et al. Interface-engineered templates for molecular spin memory devices. *Nature* **493**, 509-513 (2013).

41. K. V. Raman, Interface-assisted molecular spintronics. Interface-assisted molecular spintronics. *Appl. Phys. Rev*. **1**, 031101 (2014).

42. M. Callsen, V. Caciuc, N. Kiselev, N. Atodiresei and S. Blügel, Magnetic hardening induced by nonmagnetic organic molecules. *Phys. Rev. Lett.* **111**, 106805 (2013)

43. J. Nogués, and I. K. Schuller, Exchange bias. *J. Magn. Magn. Mater.* **192**, 203-232 (1999).

44. J. Hong, T. Leo, D. J. Smith, and A. Berkowitz, E. Enhancing exchange bias with diluted antiferromagnets. *Phys. Rev. Lett.* **96**, 117204 (2006)

45. K. Ando et al. Electrically tunable spin injector free from the impedance mismatch problem. *Nat. Mater*. **10**, 655-659 (2011).

46. J. C. Rojas-Sánchez et al. Spin Pumping and Inverse Spin Hall Effect in Platinum: The Essential Role of Spin-Memory Loss at Metallic Interfaces. *Phys. Rev. Lett.* **112**, 106602 (2014).

47. D. Sun, et al. Inverse Spin Hall Effect from pulsed Spin Current in Organic Semiconductors with Tunable Spin-Orbit Coupling. *Nat. Mater.* (2016). doi:10.1038/nmat4618.

48. K. Ando and E. Saitoh, Observation of the inverse spin Hall effect in silicon. *Nat. Commun.* **3**, 629 (2012).

49. Y. Shiomi and E. Saitoh, Paramagnetic Spin Pumping. *Phys. Rev. Lett.* **113**, 266602 (2014).

**Acknowledgments**

This work was supported by the National Science Foundation-Material Science & Engineering Center (NSF-MRSEC; DMR-1121252). The ISHE measurements were supported by the National Science Foundation (DMR- 1404634). Use of the National Synchrotron Light Source, Brookhaven National Laboratory was supported by the DOE BES (DE-AC02-98CH10886).

**Additional information**

Supplementary Information is available in the online version of the paper.

The authors declare no competing financial interests.

*Correspondence and requests for materials should be addressed to Z.V.V. (val@physics.utah.edu).

**Figure Captions**

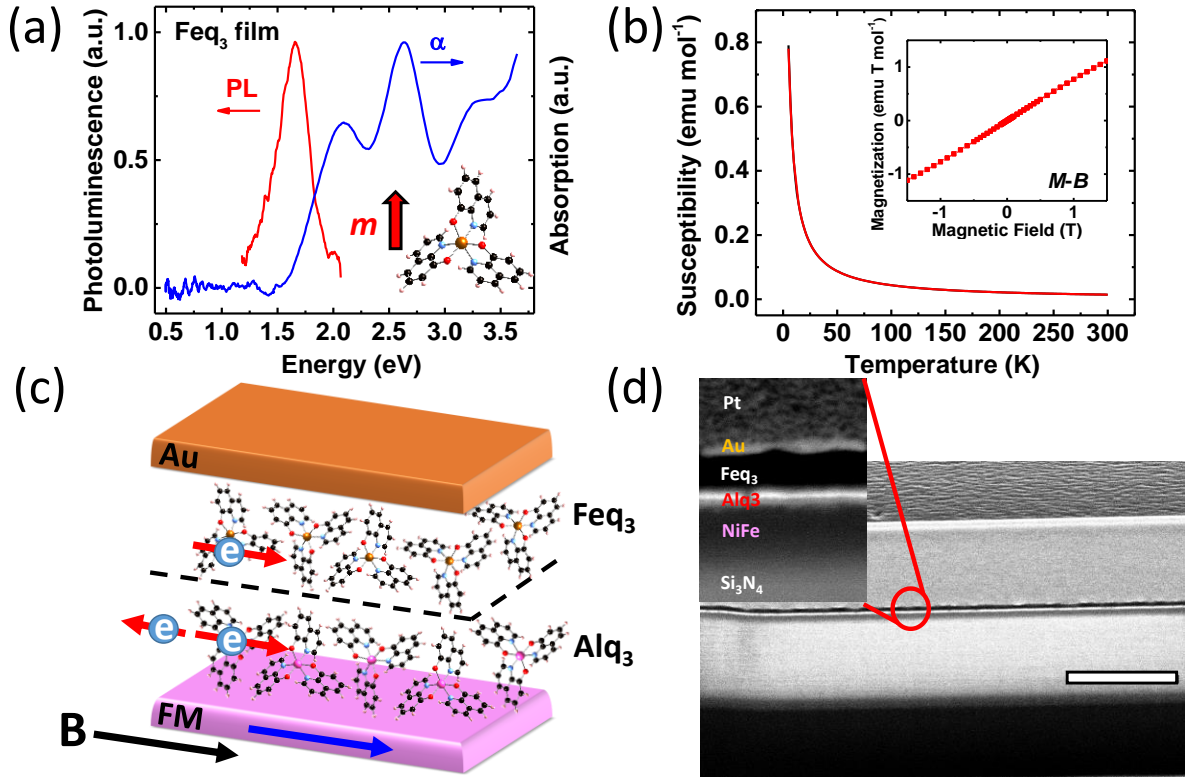

FIG. 1. (Color online) Feq$_3$ film properties and basic device characterization. (a) Absorption and photoluminescence spectra of an evaporated Feq$_3$ thin film. The inset shows the molecular structure of Feq$_3$ that contains a Fe$^{III}$ ion having spin, S=5/2 (see also S. I. Fig. S1). (b) Magnetic susceptibility of a Feq$_3$ pristine film as a function of temperature, $T$, measured by SQUID magnetometer. The inset shows $M(B)$ response characteristic of paramagnetic behaviour. (c) Schematic structure of an 'OSV-like' device that consists of a FM bottom electrode, organic spacer layer (Alq$_3$), organic spin filter layer (Feq$_3$), and capped with a nonmagnetic Au electrode. The external magnetic field ***B*** is applied parallel to the film. Spin aligned carriers of both spin orientations are injected from the FM electrode through the Alq$_3$ layer and undergo spin filtering by the Feq$_3$ layer (where one spin orientation is filtered) before reaching the Au electrode. The blue arrow indicates the magnetization direction in the FM electrode. (d) Scanning electron micrograph (SEM) of a OSV-like device cross section. The inset shows a magnified SEM image in which different layers can be clearly distinguished, as labeled. The scale bar is 1 µm.

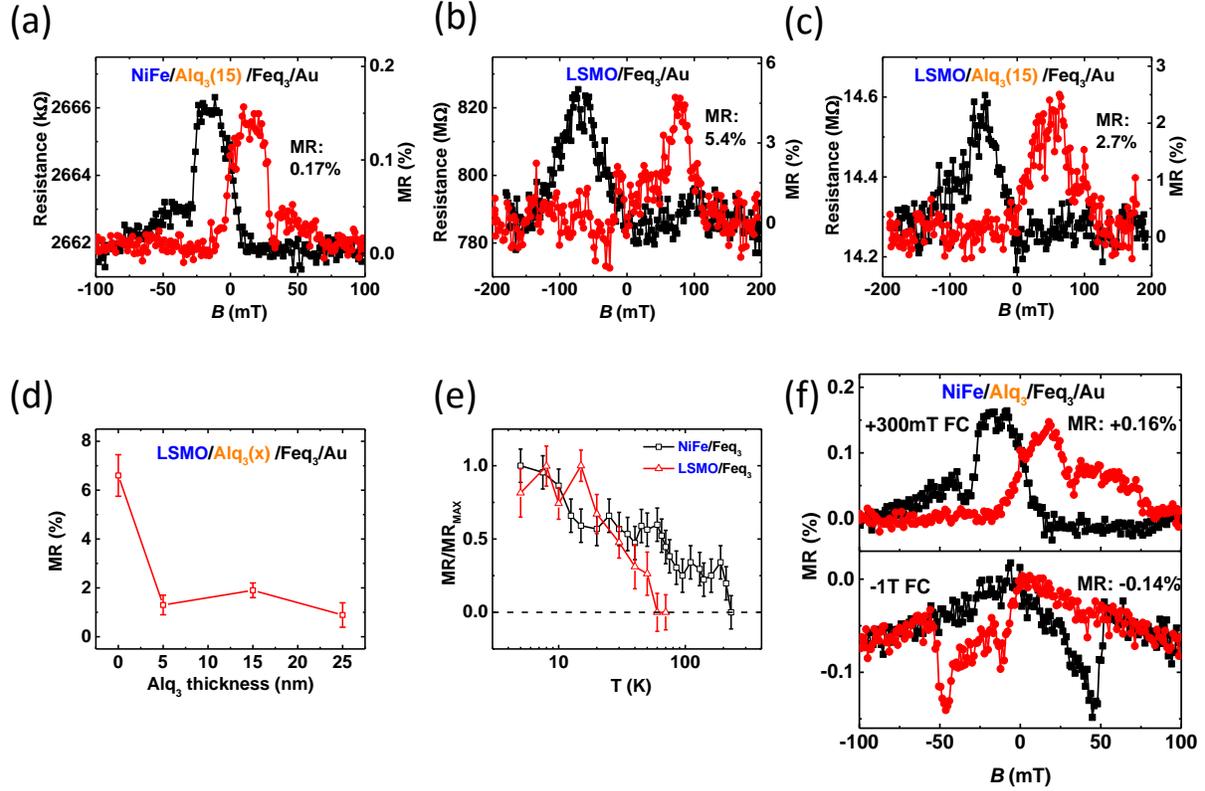

FIG. 2. (Color online) MR(*B*) response of various OSV-like devices achieved via electrical spin-injection from the FM electrode. (a) and (b), Typical MR(*B*) responses of NiFe-based and LSMO-based OSV-like devices, respectively, with the same Alq3 spacer thickness (15 nm). (c) MR(*B*) response of LSMO-based OSV-like device without the Alq3 spacer. (d) The maximum MR value, $MR_{max}$, of the LSMO-based OSV-like devices measured as a function of the Alq3 thickness, *d*. (e) $MR_{max}$ vs. temperature of NiFe and LSMO-based OSV-like devices, normalized to $MR_{max}$ at 5K. (f) MR(*B*) response of a NiFe-based OSV-like device measured at 5K, subjected to two different field cooling (FC) of 300 mT (upper panel) and -1 T (lower panel), respectively.

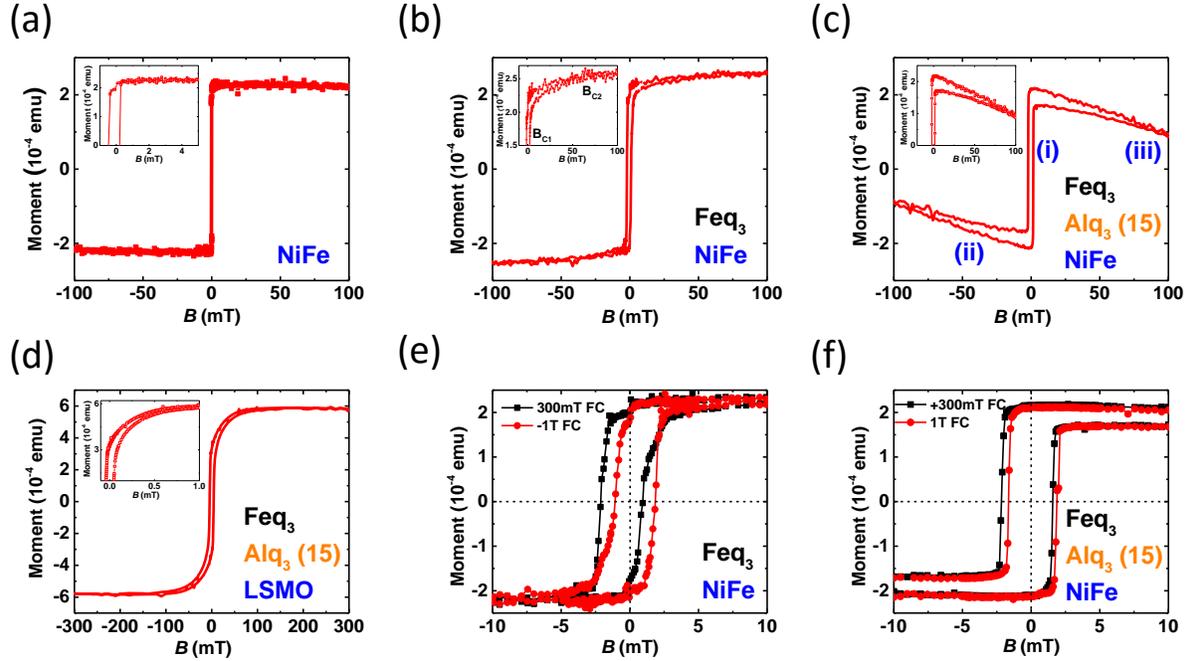

FIG. 3 (Color online) SQUID magnetometry of OSV-like device structures. (a) to (d), $M(B)$ response for NiFe, NiFe-Feq$_3$, NiFe-Alq$_3$-Feq$_3$, and LSMO-Alq$_3$-Feq$_3$ structures, respectively, plotted up to ±100 mT. The insets magnify the $M(B)$ responses that exhibit the additional hysteresis response of the deposited Feq$_3$ film onto the NiFe substrate. In panel **c**, the abrupt transition due to the FM substrate, broad transition from the Feq$_3$ layer, and the diamagnetic background response from the Alq$_3$ layer are denoted as (i), (ii) and (iii), respectively. (e) and (f), $M(B)$ responses of NiFe-Feq$_3$ and NiFe-Alq$_3$-Feq$_3$ structures, respectively, upon opposite field cooling, plotted up to ±10 mT. All $M(B)$ measurements were performed at 5K.

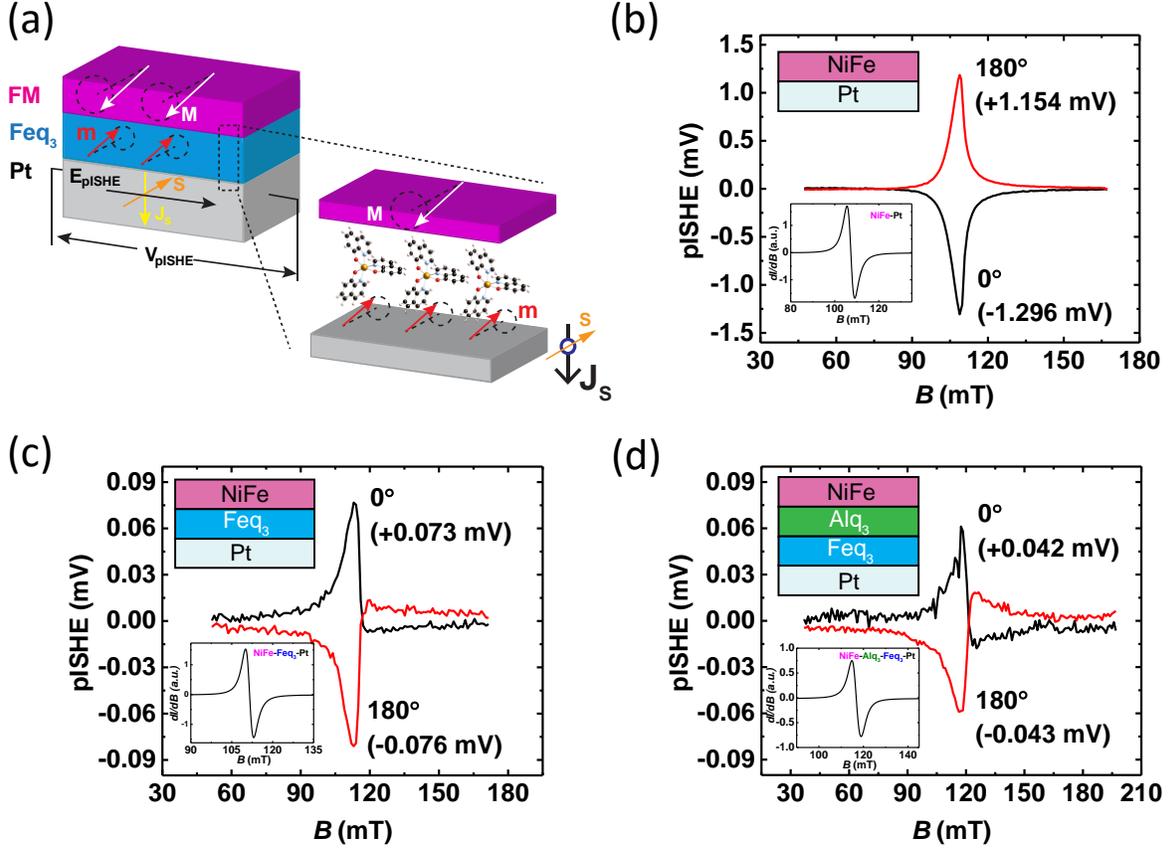

FIG. 4. (Color online) ISHE response in various Feq3-based devices generated via spin pumping. (a) Left panel: schematic illustration (not to scale) of the FM/Feq3/Pt device. $B_0$ and $M$ denote, respectively, the static external magnetic field and dynamic magnetization in the FM film that precesses about $B_0$. $J_S$, $S$, $E_{ISHE}$, and $V_{pISHE}$ denote, respectively, the flow of the pulsed spin current, spin polarization vector, generated electric field, and detected p-ISHE voltage. Right panel shows the magnetization precession of the Feq3 layer, where $m$ and derived $S$ are antiparallel to $M$, under the influence of FM layer via the AFM exchange interaction. (b), (c), and (d), $V_{pISHE}(B)$ respective response of NiFe (15 nm)/Pt (10 nm), NiFe (15 nm)/Feq3 (7 nm)/Pt (10 nm), and NiFe (15 nm)/Alq3 (15 nm)/Feq3 (7 nm)/Pt (10 nm) devices, with device structures shown in the appropriate panels. All devices are capped with a SiO2/Cu capacitor layer to suppress the anomalous Hall effect response component [46]. The black and red lines are for in-plane magnetic field $B$ (at 0º) and $-B$ (at 180º), respectively. The lower inset in each panel shows the appropriate FMR($B$) response.

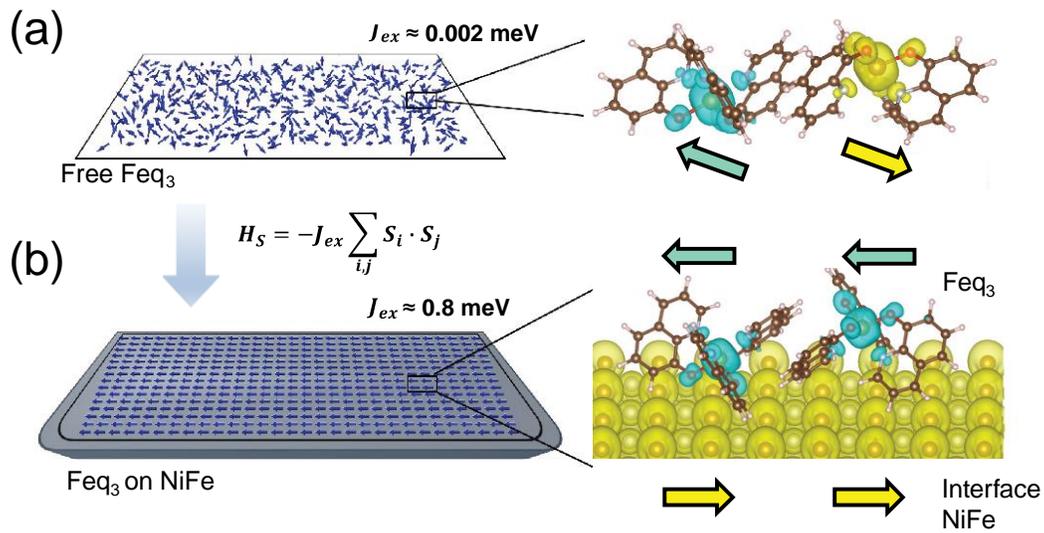

FIG. 5. (Color online) Theoretical calculations. (a) and (b), the spin density of free Feq3 molecules and Feq3 in contact with the NiFe substrate, respectively. The effective coupling constant between Feq3 molecules in two systems are labeled, deduced from Heisenberg spin lattice model. The small dark yellow and light yellow dots represent, respectively, the Fe and Ni atoms of NiFe at the interface. The yellow (blue) spheres denote spins oriented to the right (left side).

# Supplementary Information

# Spintronics Detection of Interfacial Magnetic Switching in a Paramagnetic Tris(8-hydroxyquinoline)iron(III) Thin Film

Dali Sun[1], Christopher M. Kareis[2], Kipp J. van Schooten[1], Wei Jiang[3], Marzieh Kavand[1], Royce A. Davidson[2], Gene Siegel[3], William W. Shum[2], Chuang Zhang[1], Ashutosh Tiwari[3], Christoph Boehme[1], Feng Liu[3], Peter W. Stephens,[4] Joel S. Miller[2], Z. Valy Vardeny[1],*

[1]*Department of Physics & Astronomy, University of Utah, Salt Lake City, Utah, 84112.*

[2]*Department of Chemistry, University of Utah, Salt Lake City, Utah, 84112.*

[3]*Department of Material Science & Engineering, University of Utah, Salt Lake City, Utah 84112.*

[4]*Department of Physics & Astronomy, Stony Brook University, Stony Brook, NY 11794.*

*To whom correspondence should be addressed: val@physics.utah.edu

**Section 1. Feq$_3$ structure and characterization**

The crystal structure of Feq$_3$ FIG. S1, was determined from high resolution powder diffraction patterns, FIG. S2 [S1] collected at beamline X16C at the National Synchrotron Light Source at Brookhaven National Laboratory at ambient temperature. Samples were sealed in thin-walled glass capillaries of nominal diameter 1.0 mm, which were spun during data collection. A Si(111) double-crystal monochromator selected a highly collimated incident beam of λ ~ 0.7 Å x-rays. Structures were solved and refined using TOPAS-Academic [S2] and the structure was visualized with the program VESTA [S3].

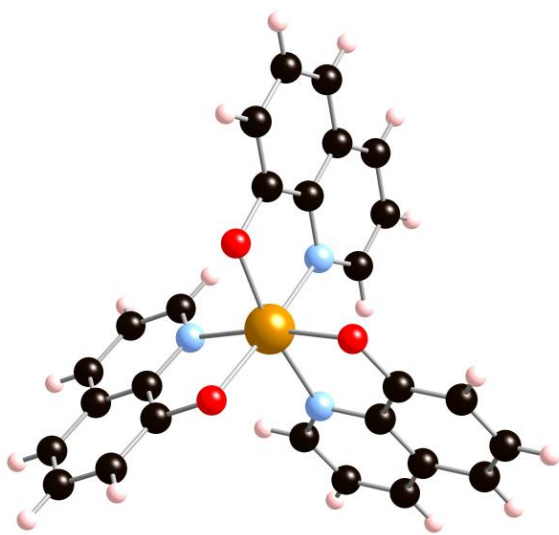

**FIG. S1. Molecular structure of Feq$_3$.** The Fe-O distances are: 1.964(15), 1.989(12), and 1.987(10) Å, and the Fe-N distances are: 2.193(9), 2.157(8), and 2.209(8) Å.

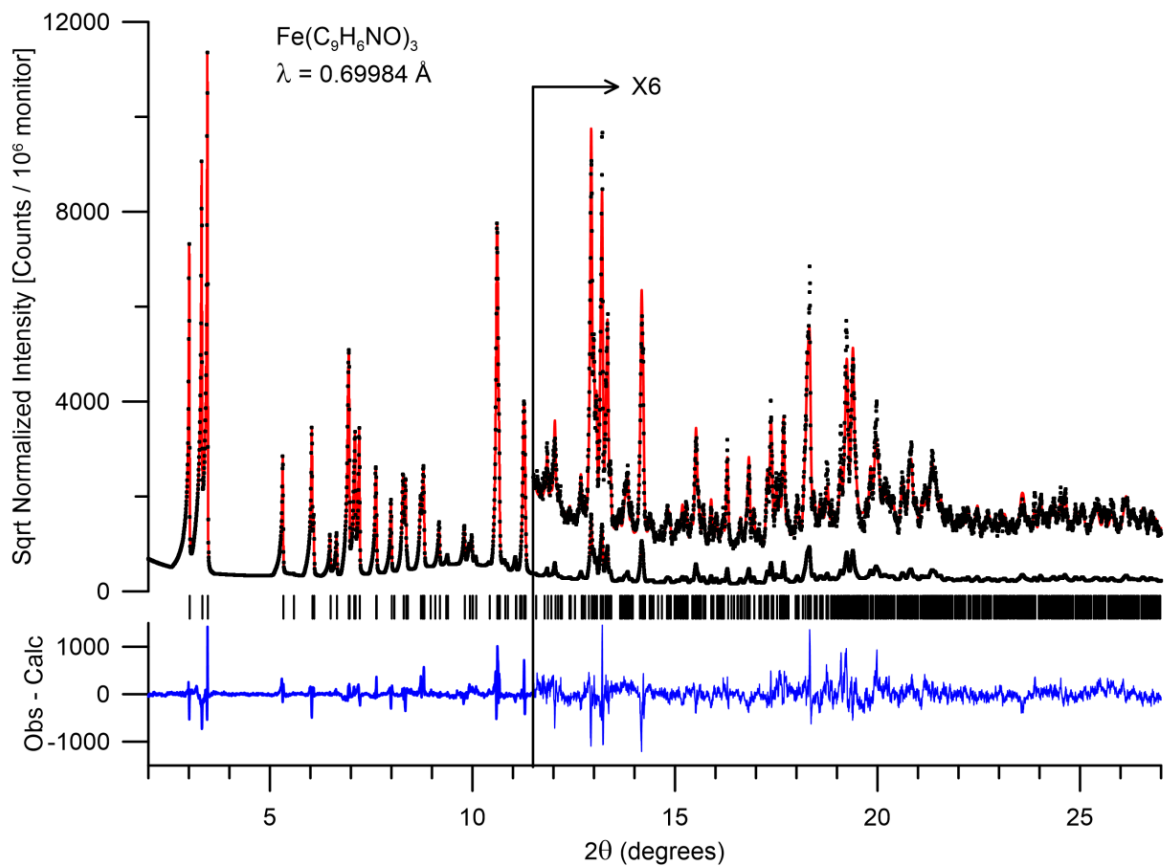

**FIG. S2**. High-resolution synchrotron powder diffraction data (dots) and Rietveld fit of the data for Feq$_3$. The lower trace is the difference, measured minus calculated, plotted to the same vertical scale.

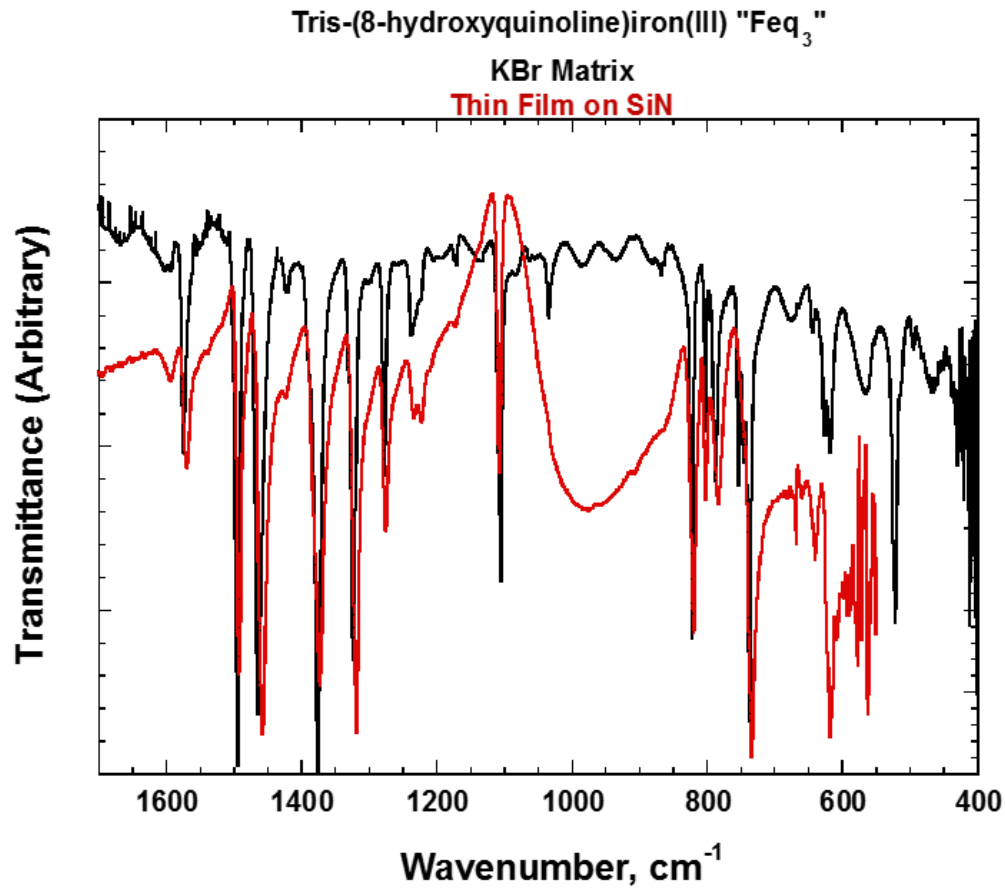

**FIG. S3. Feq₃ molecule structure and thin film characterization**. The stability of thermal evaporated Feq$_3$ thin film on Si$_3$N$_4$ substrate was investigated by Fourier transform infrared spectroscopy (FTIR). As a reference for analyzing the IR spectra, Feq$_3$ was dispersed in a KBr powder and pressed into a pellet for FTIR measurements. The absorption peaks for the two samples match each other in the energy range within 600 to 1600 cm$^{-1}$. The additional broad background between 900 to 1200 cm$^{-1}$ originates from the Si$_3$N$_4$ substrate.

**Section 2. Anisotropic Magnetoresistance (AMR) response of the electrode used in the OSV-like devices.**

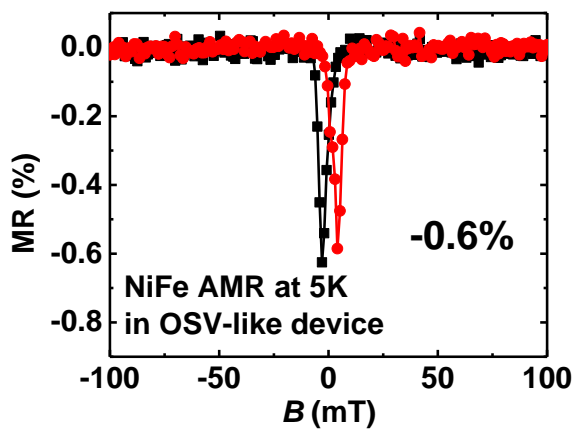

**FIG. S4. Anisotropic Magnetoresistance (AMR) response of the NiFe electrode used in the OSV-like device.** The AMR($B$) response shows the coercive field, $B_{C1}$ of the pristine NiFe electrode. The coercive field of the FM electrode in the OSV-like device is slightly increased compared to the pristine NiFe. This is attributed to magnetic hardening that originates from the exchange interaction between Feq$_3$ and FM substrate[42].

**Section 3. Intrinsic organic magnetoresistance (OMAR) in Feq₃ molecules.**

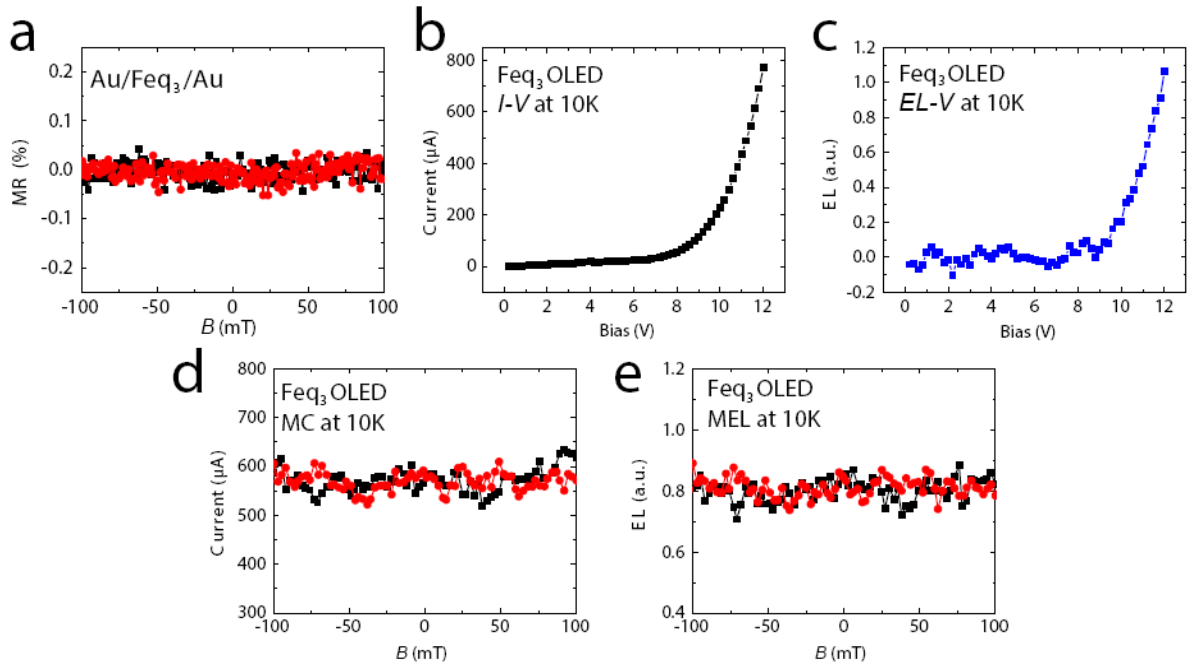

**FIG. S5. MR response of Feq3 unipolar and bipolar devices with no FM electrodes. a,** MR response for a Feq3-based device with the unipolar configuration of Au/Feq3/Au. **b** and **c**, Typical *I-V* and *EL-V* characteristic dependences of Feq3-based OLED device with the bipolar configuration of ITO/Feq3/Ca/Al. **d** and **e**, magneto-current and magneto-electroluminescence response, respectively in Feq3-based OLED device with no FM electrode. No clear organic magnetoresistance (OMAR)[34] effect from the Feq3-based device is observed.

**Section 4.** *I-V characteristics and resistance changes as function of temperature in OSV-like devices.*

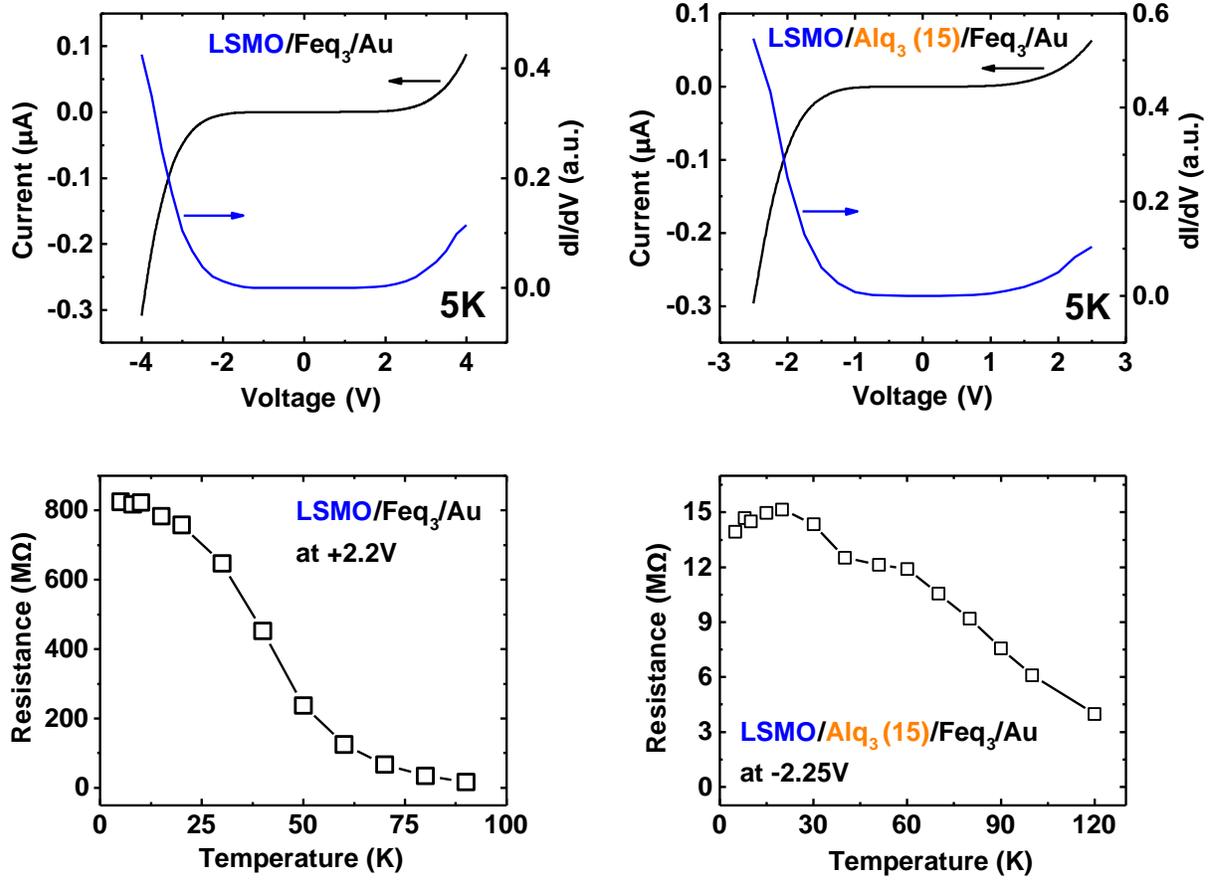

**FIG. S6.** *I-V characteristics and R vs. T in OSV-like devices.* **a** and **b,** *I-V* (black) and *dI/dV* (blue) characteristics in LSMO/Feq3/Au and LSMO/Alq3/Feq3/Au OSV-type devices, respectively. The non-linear behavior at both voltage bias indicate spin transport occurs in the diffuse regime, rather than tunneling. *dI/dV* curves show that no zero bias anomaly exists in both devices. **c** and **d**, R vs. T for two devices, respectively.

**Section 5. Non-hysteresis MR($B$) background in LSMO-based OSV-like devices.**

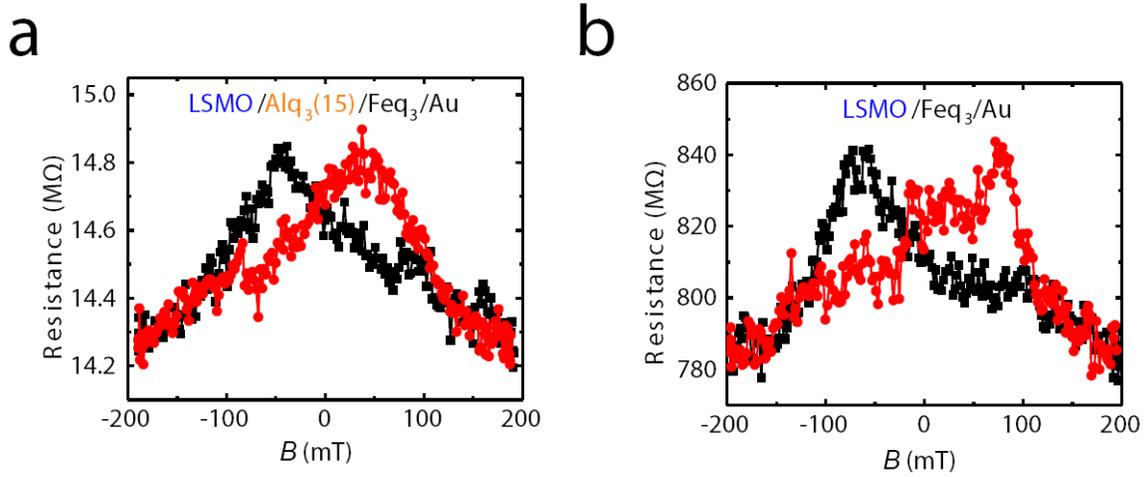

**FIG. S7. MR($B$) that contains both hysteretic and non-hysteretic responses in LSMO-based OSV-like devices. a** and **b,** As obtained MR($B$) response of the LSMO/Alq$_3$/Feq$_3$ and LSMO/Feq$_3$ OSV-type devices, respectively, that include the non-hysteretic background response. The linear –like background, non-hysteretic MR($B$) response is attributed to the MR related to the LSMO/OSEC interface[37]. Representative FIG.s in the main text had this linear response subtracted out for clarity.

**Section 6. The Feq₃ layer switching field in the OSV-like device configuration**

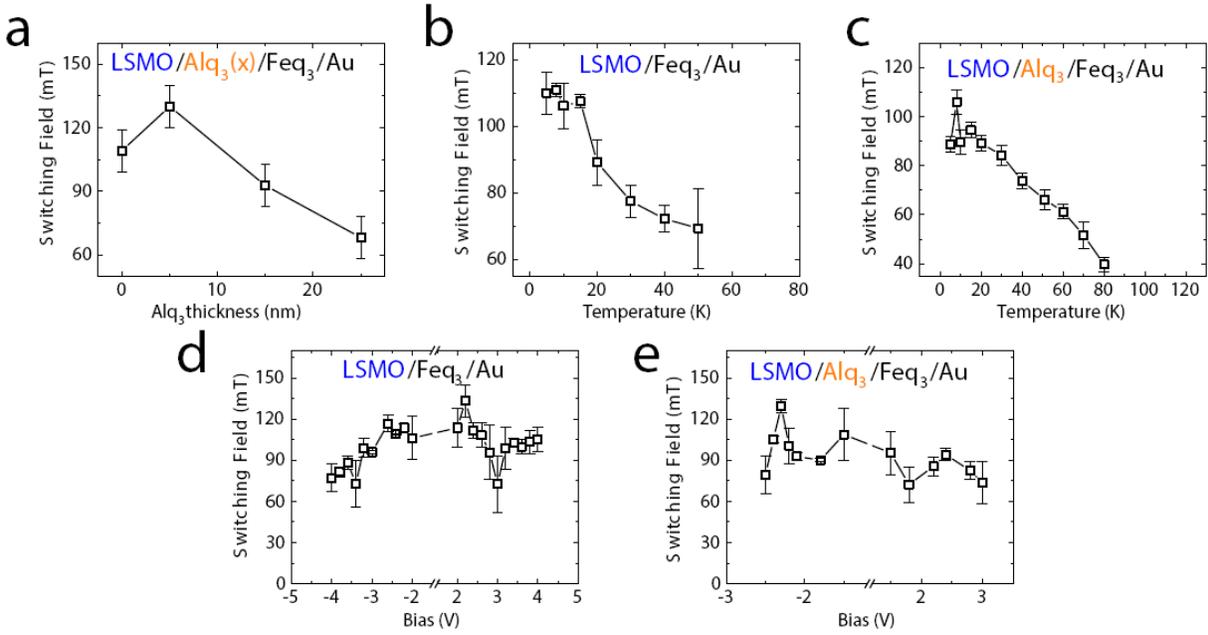

**FIG. S8. The Feq₃ switching field ($B_{C2}$) as a function of the Alq₃ spacer thickness, temperature, and bias voltage, respectively. a,** $B_{C2}$ as a function of Alq₃ thickness in LSMO-based OSV-type devices. **b** and **c,** $B_{C2}$ as a function of the device temperature in LSMO/Feq₃ and LSMO/Alq₃/Feq₃ OSV-type devices, respectively. **d** and **e**, $B_{C2}$ as a function of the bias voltage in LSMO/Feq₃ and LSMO/Alq₃/Feq₃ OSV-type devices, respectively. No clear change in the $B_{C2}$ is observed. Thus, a possible effect from current-induced-magnetic switching in the Feq₃ layer may be excluded.

## Section 7. Temperature dependence of MR(B) response in OSV-like device

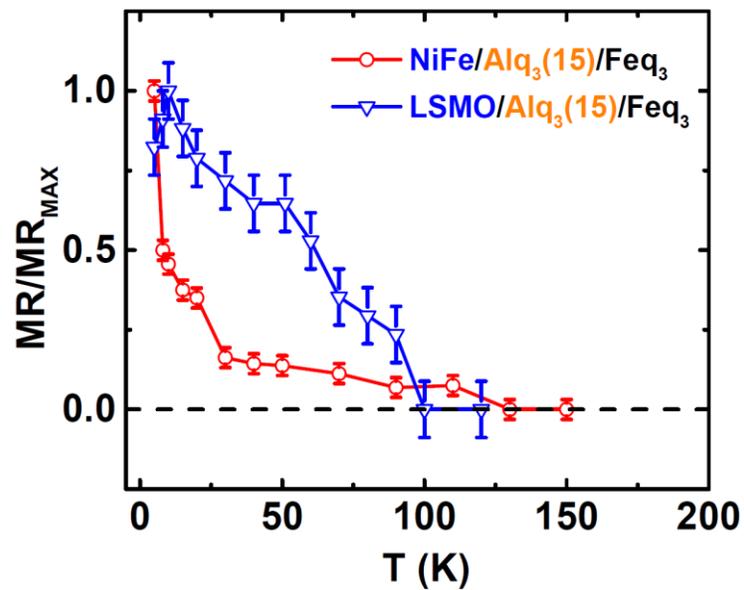

**FIG. S9.** MR$_{max}$ vs. temperature of various OSV-like devices (with Alq$_3$ spacer), normalized to MR$_{max}$ at 5K.

**Section 8. The bias voltage dependence of MR(*B*) response in OSV-like device**

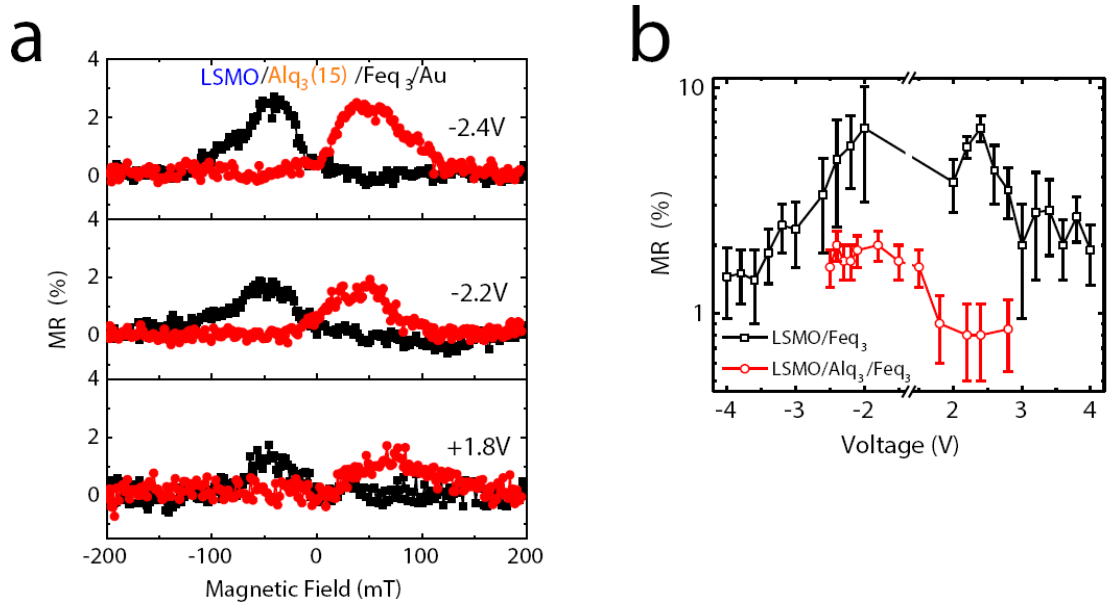

**FIG. S10. Bias dependence of MR response. a,** MR(*B*) response and **b,** $MR_{max}$ dependencies on the bias voltage in LSMO-based OSV-like device measured at 5K. The *V*-dependence of the MR(*B*) response in the OSV-like devices is similar to that of conventional LSMO-based OSV devices in the literature[4-9].

**Section 9. MR polarity reversal in NiFe/Feq$_3$ and LSMO/Feq$_3$ OSV-like devices**

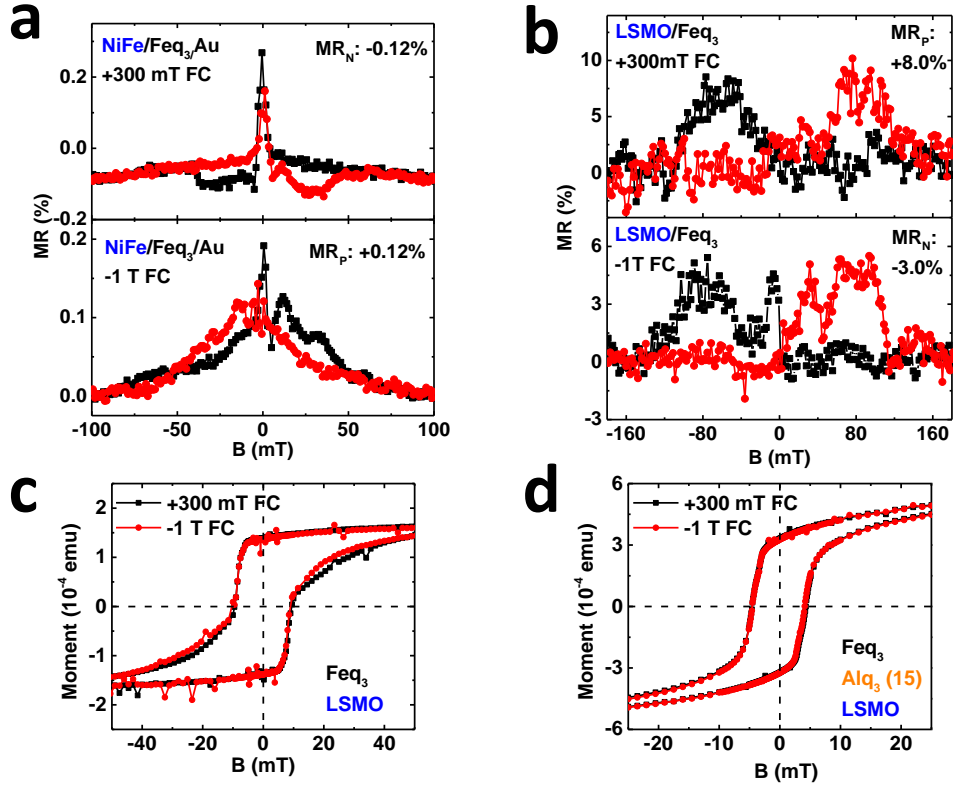

**FIG. S11. MR polarity reversal in NiFe/Feq$_3$ and LSMO/Feq$_3$ devices and *M(H)* response upon different field cooling. a** to **b,** observation of MR(*B*) polarity reversal in NiFe/Feq$_3$ and LSMO/Feq$_3$ device, respectively. The observed peaks in the NiFe/Feq$_3$ device at *B*~0 indicate TAMR(*B*) response from the NiFe electrode that can be easily separated from the transport-related MR(*B*) response. The MR reversal in LSMO/Feq$_3$ device is incomplete (both signs exist) probably due to relative weak AFM coupling between LSMO and Feq$_3$. It also supports our assumption that the MR reversal is caused by noncollinear spin orientation in Feq$_3$ layer, similar as in Ref. S5 and S6. This weak coupling is reduced after the Alq$_3$ layer is inserted, and consequently no MR reversal is observed in LSMO/Alq$_3$/Feq$_3$ device. **c** and **d,** *M(B)* responses of LSMO-Feq$_3$ and LSMO-Alq$_3$-Feq$_3$ structures upon opposite field cooling. While LSMO-Feq$_3$ still presents a slight exchange bias behavior upon opposite field cooling, the shift of *M(B)* in LSMO-Alq$_3$-Feq$_3$ is barely detectable, which is consistent with the no MR polarity reversal observed in the devices.

**Section 10. p-ISHE response in NiFe/Pt and NiFe/Feq₃/Pt structures measured via pulsed MW spin-pumping**

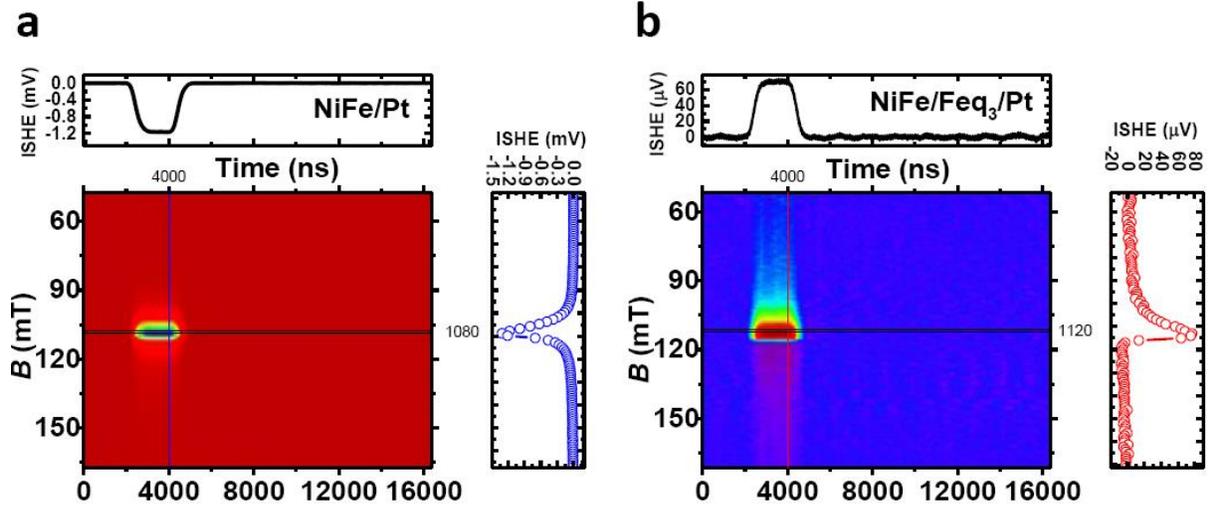

**FIG. S12. a,** Time (**t**) and field (**B**) responses of p-ISHE voltage in a prototype NiFe/Pt ISHE-type device under 1 kW microwave excitation. The black solid line in the top panel indicates the moving-average ISHE voltage response for two devices. The blue/red data points in the right panel shows the field dependence of p-ISHE response at **t** = 4 μs. The colour plot shows a resonance at $B=B_{res}$=108 mT. **b,** the same as in a, but for the NiFe/Feq₃/Pt device, where the FMR resonance field, $B_{res}$ shifts to 112 mT. Note the reversed ISHE polarity in this device, respect to the NiFe/Pt device.

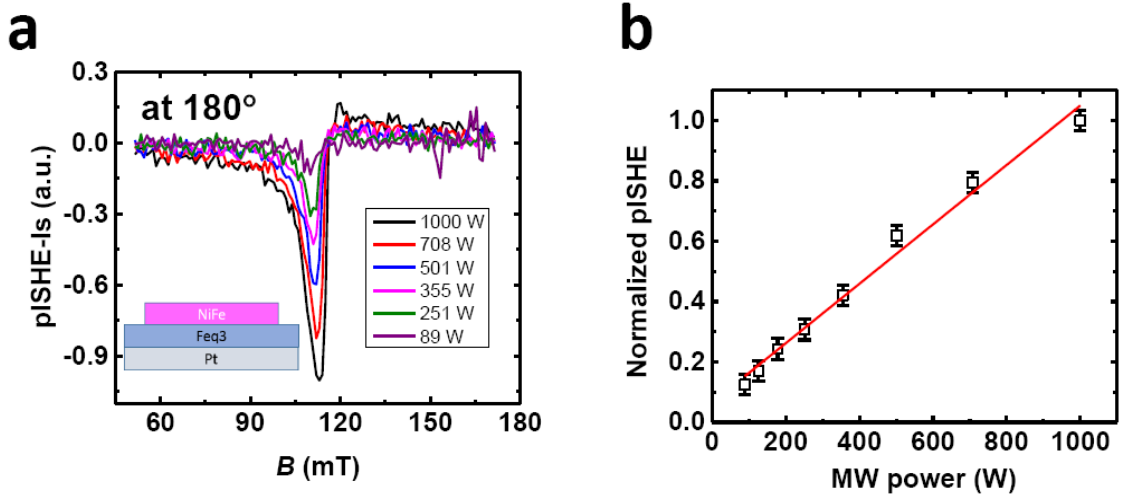

**FIG. S13.** MW power dependence response of p-ISHE response in NiFe/Feq3/Pt ISHE-type device. **a,** field dependence of p-ISHE(B) response measured at $\theta_B=180°$ and different MW powers. **b,** normalized $V_{pISHE}$ magnitude as function of the MW power (open squares). The red line through the data points is a linear fit.

We measured the pulsed-ISHE responses in a bilayers with smaller Feq3 thickness (~5 nm); see FIG. S14 below. The p-ISHE polarity in this device is still reversed as compared to the NiFe/Pt device. In addition the p-ISHE response is indeed larger (~93 µV) due to the enhanced exchange coupling at smaller Feq3 thickness.

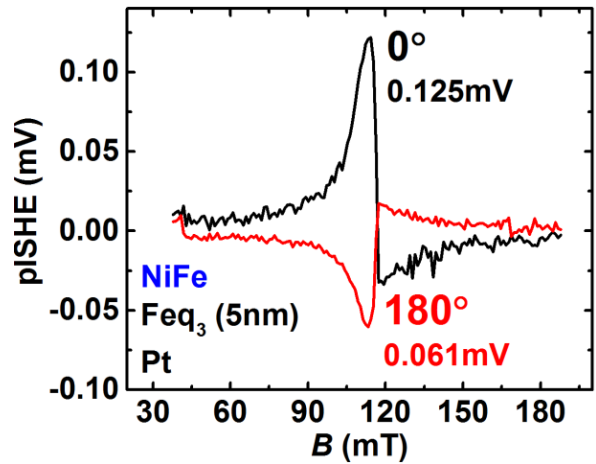

**FIG. S14. Pulsed-ISHE(*B*) response in NiFe/Feq3(5nm)/Pt device.** The ISHE value is extracted from the relation V(pISHE)=(V(0°)-V(180°))/2.

**Section 11.1 Calculation of the exchange coupling constant in a single layer Feq$_3$**

First-principles calculations were carried out using local spin density approximation (LSDA) with onsite Coulomb interactions and projector augmented-wave method in Vienna ab-initio simulation package (VASP) [S7,S8] based on density functional theory, in which an additional on-site Hubbard-U term is included on the iron(III) (U=6.0 eV, J=0.9 eV). The U values were tested to give the 5/2 spin state for Feq$_3$, in agreement with the experiment. A cut-off energy is chosen at 400 eV; a K-point mesh of 3×3×1 is used for Brillouin zone sampling; the energy convergence criteria is set at 0.1 meV.

The Heisenberg spin lattice model was used to match the first-principles calculation results in order to derive the exchange constant, $J_{ex}$. Consider a 2D spin lattice of 5/2 spins composed of Feq$_3$ molecules; the lattice Hamiltonian is defined as: $\mathcal{H}_S = -J_{ex} \sum_{i,j} \boldsymbol{S_i} \cdot \boldsymbol{S_j}$, where $J_{ex}$ is the 'effective' exchange coupling constant, $S_{i(j)}$ is the spin momenta of different Feq$_3$ molecules. Only nearest-neighbor (NN) exchange interactions were considered, the energy difference between different spin configurations, such as FM vs. AFM can then be calculated by lattice summation. For example, in a freestanding monolayer Feq$_3$ film, each Feq$_3$ has three NNs, the energy per unit cell for the FM and AFM configuration is -75/4 $J_{ex}$ and +75/4 $J_{ex}$, respectively. Thus the energy difference between the FM and AFM configuration is 75/2 $J_{ex}$, which are in accord with the results from first-principles calculation to derive the $J_{ex}$ for the freestanding monolayer Feq$_3$ film.

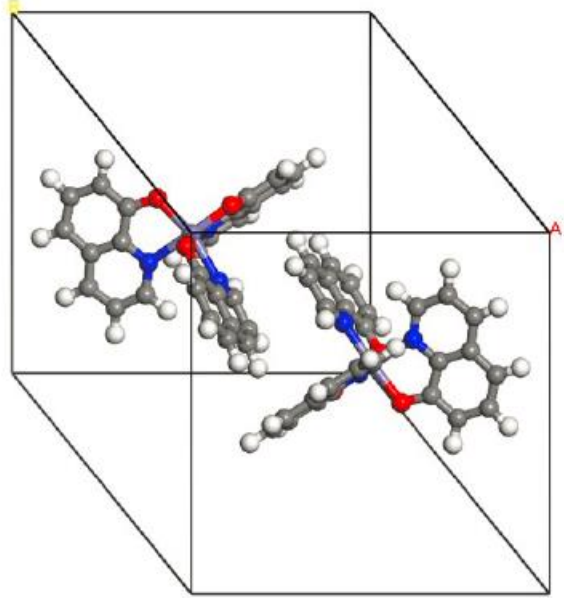

| Initial direction | $M_s (\mu_B)_{tot}$ | $M_s (\mu_B)$ | | $\Delta E (meV)$ |
| --- | --- | --- | --- | --- |
| Fe$_1$   Fe$_2$ | | Fe$_1$ | Fe$_2$ | |
| A ↑   ↓ | 0 | 5 | -5 | 0.07±0.1 |
| B ↑   ↑ | 10 | 5 | 5 | 0 |

**FIG. S15. A super cell of Feq$_3$ molecular layer.** The inset Table shows a computed energy difference between FM and AFM configuration within the Feq$_3$ layer. The second (third) column shows the total (individual) magnetic moment of Feq$_3$ molecules. By mapping the Heisenberg spin lattice model to first-principles energy difference (see discussion and Methods section in the text), a negligible exchange coupling constant, $J_{ex} \sim 0.002$ meV is deduced, indicating paramagnetic response of freestanding Feq$_3$ layer.

**Section 11.2 Calculation of Substrate Induced Magnetic-ordering in Feq₃ and AFM coupling in NiFe/Feq₃ system.**

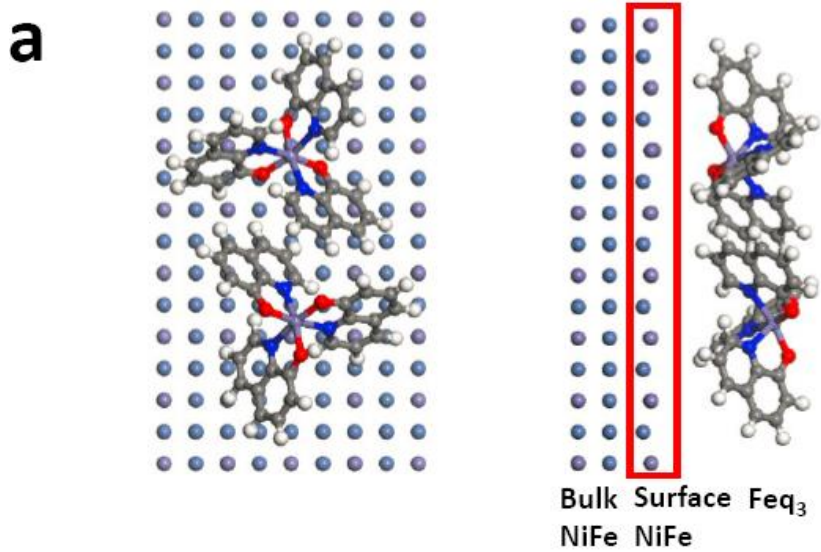

| Initial direction | | | $M_s (\mu_B)_{tot}$ | $M_s (\mu_B)$ | | | $\Delta E(meV)$ |
|---|---|---|---|---|---|---|---|
| NiFe | Fe₁ | Fe₂ | | NiFe | Fe₁ | Fe₂ | |
| A ↑ | ↑ | ↑ | 245 + 10 | 245 | 5 | 5 | 27.0±0.1 |
| B ↑ | ↑ | ↓ | 245 | 245 | 5 | -5 | 34.6±0.1 |
| C ↑ | ↓ | ↓ | 245 − 10 | 245 | -5 | -5 | 0 |

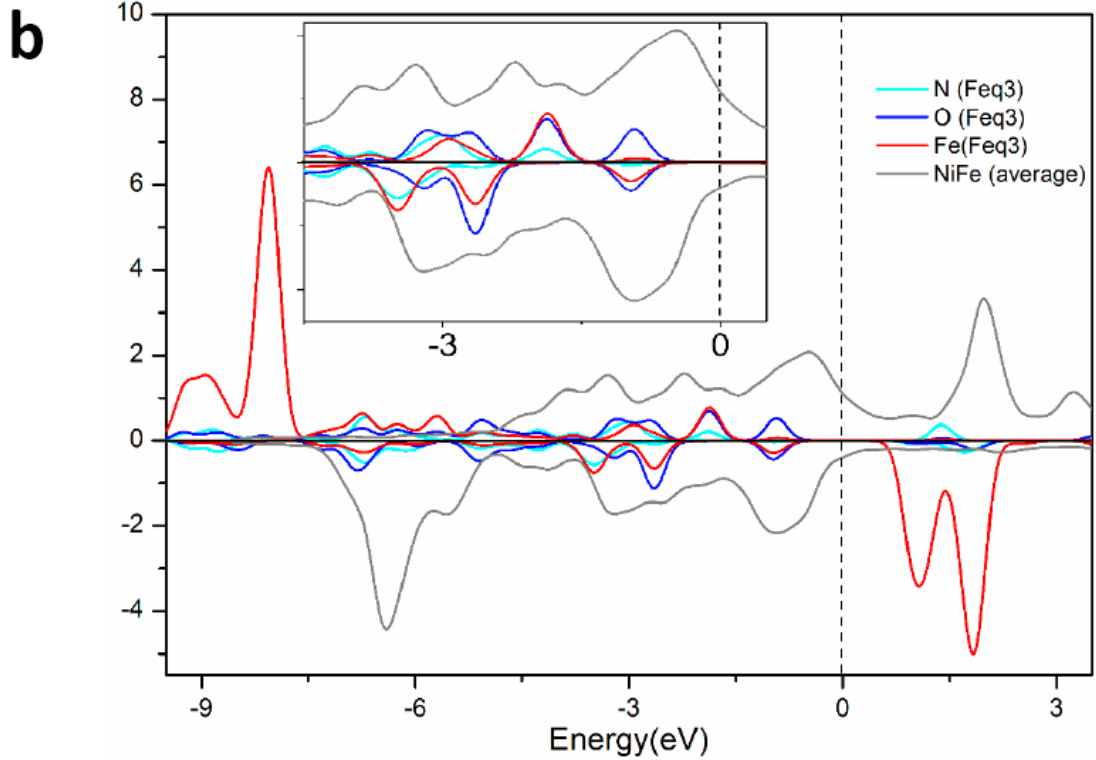

**FIG. S16. Modeling and Results of a DFT calculation of Feq$_3$ molecules on NiFe FM substrate**. **a**, Top and side views of the Feq$_3$ unit cell lying on the NiFe substrate. The Table shows the calculated energy difference between FM (state 'A') and AFM (state 'C') interface coupling between Feq$_3$ and NiFe that indicates a preferred AFM coupling ($E_{AFM}$ - $E_{FM}$ ~ −27 meV). The second (third) column in the table shows the total (individual) magnetic moment of NiFe and Feq$_3$ molecules. The coupling within Feq$_3$ molecules $J_{ex}$ is about 0.844 meV. The coupling strength between NiFe and Feq$_3$ in NiFe/Feq$_3$ is $J_{ex}'$ = -0.534 meV. **b**, Spin-resolved, partial DOS in NiFe/Feq$_3$ system, indicating Fe-O, Ni-O and Fe-N, Ni-N hybridizations [23].

**Section 11.3 Calculation of AFM coupling in NiFe/Alq$_3$/Feq$_3$ system**

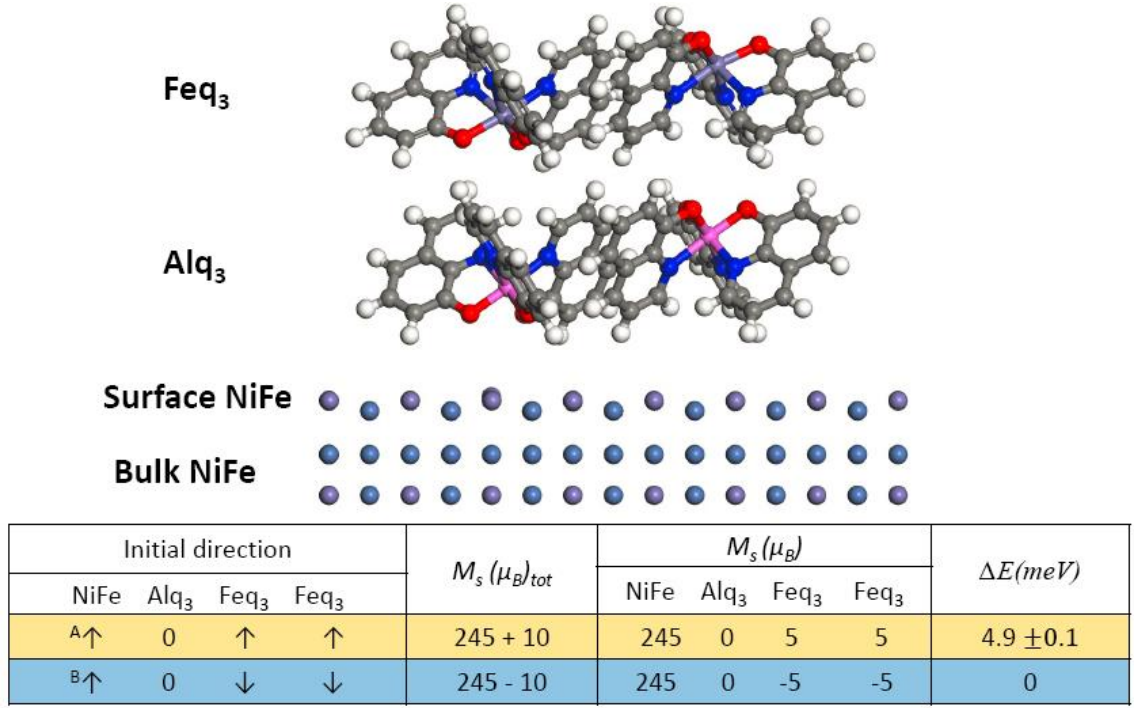

| Initial direction | | | | $M_s (\mu_B)_{tot}$ | $M_s (\mu_B)$ | | | | $\Delta E(meV)$ |
|---|---|---|---|---|---|---|---|---|---|
| NiFe | Alq$_3$ | Feq$_3$ | Feq$_3$ | | NiFe | Alq$_3$ | Feq$_3$ | Feq$_3$ | |
| A ↑ | 0 | ↑ | ↑ | 245 + 10 | 245 | 0 | 5 | 5 | 4.9 ±0.1 |
| B ↑ | 0 | ↓ | ↓ | 245 - 10 | 245 | 0 | -5 | -5 | 0 |

**FIG. S17. Substrate induced exchange interaction between NiFe and Feq$_3$ separated by an Alq$_3$ spacer layer.** Side view of Feq$_3$ and Alq$_3$ molecules sitting on the NiFe substrate. The Table inset shows the calculated energy difference between FM (state 'A') and AFM (state 'B') coupling in the NiFe/Alq$_3$/Feq$_3$ system that indicates AFM coupling to be preferred ($E_{AFM}$ - $E_{FM}$ = -4.9 meV). The second (third) column in the table shows the total (individual) magnetic moment of NiFe, Alq$_3$ and Feq$_3$ molecules. The coupling within Feq$_3$ molecules $J_{ex}$ is about 0.100 meV The coupling strength between NiFe and Feq$_3$ in NiFe/Alq$_3$/Feq$_3$ is $J_{ex}'$ = -0.097 meV, which is smaller than that of direct deposition of Feq$_3$ on NiFe (see Fig. S16a); in agreement with the observed magnetization and MR measurements.

**Section 11.4 Calculated DOS which illustrates spin-filtering of the Feq₃ layer.**

The DOS of the Feq₃ layer exhibits a clear *asymmetry* between the majority and minority spins (S. I. Fig. S18) that shows the origin of its spin-filtering capability. In addition, the energy difference, $\Delta E$ between FM and AFM coupling among neighboring molecules is very small when the Feq₃ is unpolarized, consistent with their paramagnetic response (S. I. Fig. S15).

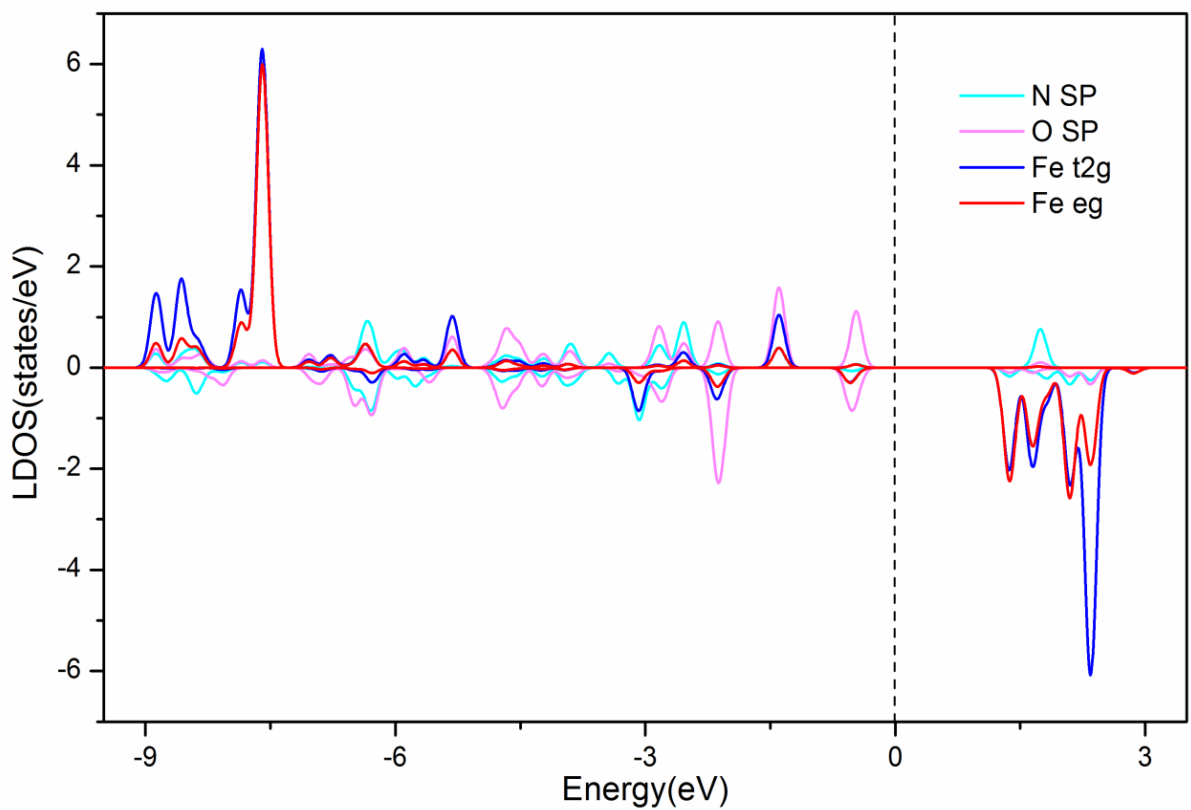

**FIG. S18. Spin-resolved, partial DOS of pristine Feq₃ molecules**. The spin filtering effect that originates from the spin-polarized DOS of Feq₃ is caused by strong exchange splitting between spin up and spin down states. The Fe atom contributes the most of the spin filtering due to the substantial gap between the spin up and spin down energy levels. The Fermi level is set at zero energy.

**Section 11.5 Model for the obtained MR(*B*) response in OSV-like devices based on Feq₃ spin filtering**

We propose a phenomenological model for explaining the MR(*B*) response in the OSV-like devices based on the Feq₃ molecules. The model is based on an AFM exchange interaction between the FM substrate the Feq₃ in the deposited OSEC layer, as revealed by the magnetization and ISHE measurements. Since this interaction decays with the distance, *r* between the two films, we envision two types of Feq₃ 'sublayers' within the deposited OSEC film which are formed upon the application of an external magnetic field, **B** (Fig. S19). These are: sublayer A (B) that is closer (farther) to the FM substrate where the molecular spins have (do not have) substantial AFM interaction with the magnetic substrate. We note, however that only sublayer A exists in the Feq₃ film grown for the ISHE measurements, since the thickness is only 7 nm, too thin to accommodate two sublayers, A and B.

The OSV-like MR(*B*) response occurs as following. When the OSV-like device is cooled in a 'negative' **B**, and the sweeping field starts at B<0, then the FM electrode magnetization aligns parallel to the field direction. In this case, due to the AFM interaction with the substrate, the Feq₃ spins in sublayer A are aligned antiparallel to the FM magnetization direction. However because of lack of AFM interaction with the substrate, the spin orientation of the Feq₃ molecules in sublayer B are aligned in the direction of the applied field (i.e. negative, Fig. S19 insets). The two Feq₃ sublayers maintain their mutual relative antiparallel magnetization alignment via a weak magnetic interaction, and the device resistance is in a 'low resistance' state because the thicker sublayer B acts as a spin filter due to its spin alignment along **B**. At $B_{C1} < B < B_{C2}$, the FM electrode switches its magnetization direction and, due to the AFM interaction the spins in Feq₃ sublayer A also switch their alignment. At this stage the spin alignment in the two Feq₃ sublayers are parallel to each other and antiparallel with **B** and **M** in the FM substrate; consequently the device is in a 'high resistance' state. This intermediate 'high resistance' state switches back to a 'low resistance' state at $B>B_{C2}$ when the external field overcomes the coupling strength between the two Feq₃ sublayers, A and B. This results in a spin valve-like MR(*B*) response that we measured.

Our proposed model contains three key assumptions: (i) a substrate-induced magnetic ordering in Feq₃ sublayer A that prefers an AFM coupling with the FM electrode, which may also be mediated through several layers of Alq3; (ii) two magnetically different and coupled Feq3 sublayers (A and B, respectively), of which spin orientations depend on the magnetic field cooling history (S.I. Fig. S20); and (iii) spin-filtering capability of the paramagnetic Feq3 sublayer of which induced magnetic order is governed by (i) and (ii) (S.I. Fig. S18). The first assumption was already experimentally demonstrated for a different paramagnetic molecular monolayer of Fe-porphyrin evaporated on a FM substrate [23,25], where an indirect, super-exchange interaction occurs between the $Fe^{II}$ ions and the FM substrate that induces magnetic order in the paramagnetic molecules. This is verified for the Feq3 here via our detailed magnetization and ISHE measurements. The second and third assumptions are also confirmed by theoretical calculation in Fig. S20 and Fig. S18, respectively.

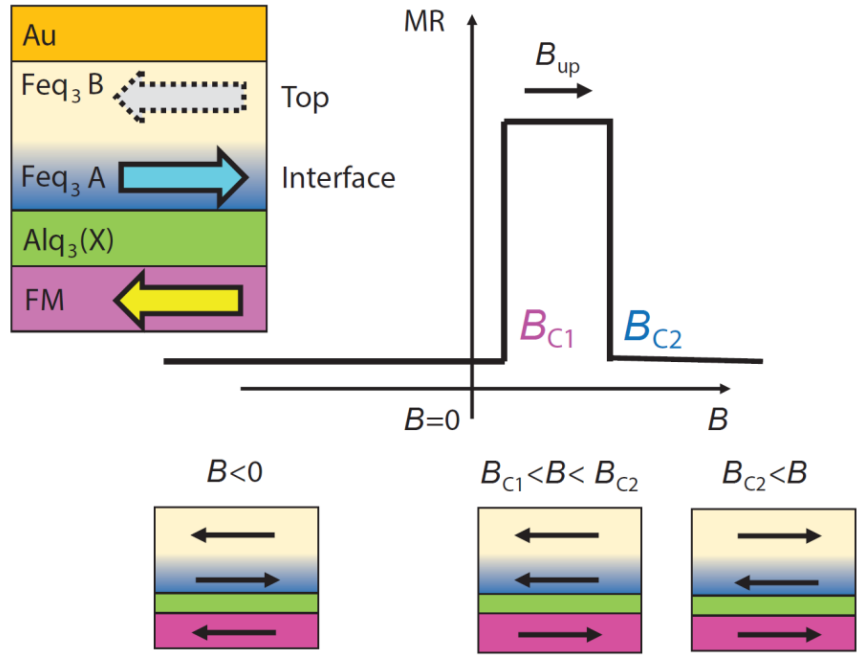

**FIG. S19. Model for the spintronics device operation.** Schematic OSV-like device that shows the magnetization directions for the bottom FM electrode, Feq3 interface, and top sublayers, respectively, when sweeping the external magnetic field $B$ in the forward direction (black arrow). Note that the sublayer A (blue) gradually transforms to sublayer B (light yellow) for Feq3 away from FM electrode, as indicated by graduate color. The left section shows the magnetization directions for the various OSV-like layers upon cooling, before sweeping the field upward. The bottom three parts show the magnetization directions upon sweeping the field through the FM coercive field, $B_{C1}$, (middle section) and the switching field, $B_{C2}$, (right section).

**Section 11.6 Calculation of spin alignment between the two Feq₃ sublayers**

Once the Feq₃ spins in sublayer A are polarized by the substrate-induced AF exchange interaction, then the spins in Feq₃ sublayer B become pinned (see Fig. S20). Consequently, the energy difference between parallel and antiparallel spin configurations of Feq₃ sublayer B is larger, and thus spin alignment is preferred. This explains the hysteretic FM switching in the Feq₃ multi-layers, as obtained in our MR and magnetization measurements. Furthermore, this interlayer coupling is relatively weak and a non-collinear spin orientation can form that can be easily reversed in different field directions [S5,S6]. This is also consistent with our observations (S. I. Fig. S20).

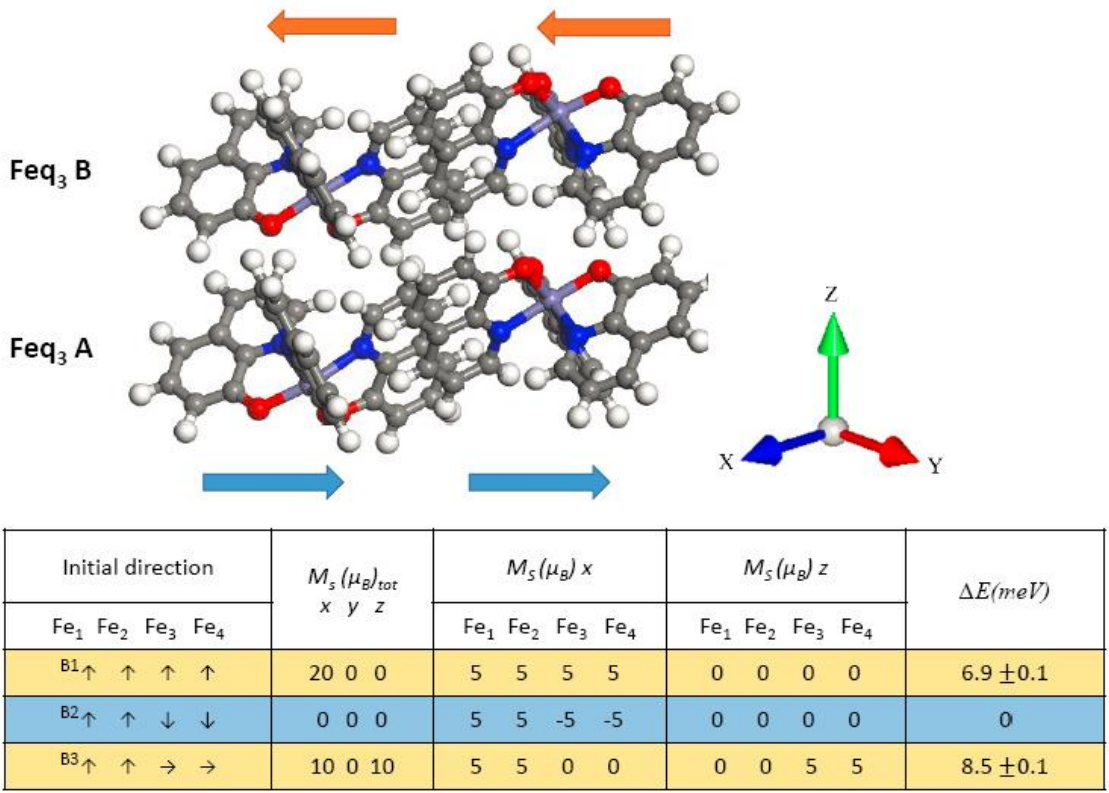

| Initial direction Fe₁ Fe₂ Fe₃ Fe₄ | $M_s (\mu_B)_{tot}$ x y z | $M_s (\mu_B)$ x Fe₁ Fe₂ Fe₃ Fe₄ | $M_s (\mu_B)$ z Fe₁ Fe₂ Fe₃ Fe₄ | $\Delta E (meV)$ |
|---|---|---|---|---|
| B1 ↑ ↑ ↑ ↑ | 20 0 0 | 5 5 5 5 | 0 0 0 0 | 6.9 ±0.1 |
| B2 ↑ ↑ ↓ ↓ | 0 0 0 | 5 5 -5 -5 | 0 0 0 0 | 0 |
| B3 ↑ ↑ → → | 10 0 10 | 5 5 0 0 | 0 0 5 5 | 8.5 ±0.1 |

**FIG. S20. Preferred spin alignment for the two adjacent Feq₃ sublayers.** Side view of DFT model for the two Feq₃ interface sublayers. The spin orientations of the top (B) and interface (A) Feq₃ sublayers in the model used to explain the OSV-like device operation. The Feq₃ (A) sublayer is closer to the NiFe FM electrode and its magnetization is aligned antiparallel to the electrode magnetization direction (blue arrows with solid outline). The Feq₃ (B) sublayer is influenced by the polarized interface sublayer and shows spin order (grey arrows with dot

outline) due to a weak exchange coupling with the top sublayer. For simplicity only two unit cells of Feq$_3$ are considered (one unit cell contains two Feq$_3$ molecules [S4]). The inset Table shows the calculated energy difference between different spin orientations in the Feq$_3$(A)/Feq$_3$(B) sublayers. The second column in the table shows the total magnetic moment of Feq$_3$ molecules along the axis of *x*, *y*, *z*. The third and fourth column show the individual magnetic moment of Feq$_3$ molecules along '*x*' and '*z*' direction. State 'B3' in the Table shows the non-collinear switching [S5,S6] of the top Feq$_3$ layer that explains the MR polarity change in the OSV-type device upon magnetic field cooling history. State 'B2' in the Table is found to be the lowest energy state indicating that AFM coupling between the two sublayers A and B is preferred. Using the same approach as in Fig. S16, we obtained an AFM coupling constant of $J_{ex}$' = ~0.023 meV which is comparable to the coupling between Feq$_3$ and NiFe (see Fig. S16a).

**References:**

S1. Mn$^{II}$q$_3$: C$_{27}$H$_{18}$FeN$_3$O$_3$, Space group = $P\bar{1}$ ; $a$ = 6.21368(16) Å; $b$ = 13.2666(3) Å; $c$ = 14.5201(5) Å; $\alpha$ = 65.996(3); $\beta$ = 88.357(3); $\gamma$ = 83.516(2); $V$ = 1086.25(6) Å$^3$; $T$ = 298 K; $Z$ = 2; $R_{wp}$ = 0.0726; $\chi^2$ =6.83. CCDC #1045846.

S2. Coelho, A. A. *TOPAS-Acadmic* available at http://www.topas-academic.net.

S3. Momma, K. and Izumi, F. A three-dimensional visualization system for electronic and structural analysis. *J. Appl. Crystallogr.* **41**, 653 (2008)

S4. Bo Li, J.P. Zhang, X.Y. Zhang, J.M. Tian, G.L. Huang, The Correlation between Magnetism and Structures for New Solvent Free Mq$_3$ Complexes (M = Fe$^{3+}$ and Cr$^{3+}$; q = 8-quinolinato). *Inorg. Chim. Acta.* **366**, 241-246 (2011)

S5. Xiang, G., Sheu, B. L., Zhu, M., Schiffer, P. and Samarth, N. Noncollinear spin valve effect in ferromagnetic semiconductor trilayers. *Phys. Rev. B* **76**, 035324 (2007).

S6. Yoo, T. et al. Tunneling magnetoresistance from non-collinear alignment of magnetization in Fe/GaAlAs/GaMnAs magnetic tunnel junctions. *Appl. Phys. Lett.* **102**, 212404 (2013).

S7. P. E. Blöchl, Projector augmented-wave method, *Phys. Rev. B* **50**, 17953 (1994).

S8. G. Kresse and J. Joubert, From ultrasoft pseudopotentials to the projector augmented wave method, *Phys. Rev. B* **59**, 1758 (1999).